\documentclass[a4paper,10pt]{article}

\usepackage[final]{graphicx}
\usepackage{times}
\usepackage{amsmath}
\usepackage{amssymb}
\usepackage{amsthm}
\usepackage{mathtools}
\usepackage{graphicx} 
\usepackage{epsfig}
\usepackage{ifthen}
\usepackage[utf8]{inputenc}
\usepackage[english,russian]{babel}
\usepackage[T2A]{fontenc}
\usepackage{setspace}
\usepackage{url}
\usepackage[process=auto,crossref=false]{pstool}
\usepackage[svgnames]{xcolor}
\usepackage[colorlinks,allcolors=DarkCyan,unicode,pdftitle={Non-stationary elastic wave scattering}]{hyperref}
\usepackage{stackrel}
\usepackage{afterpage}
\usepackage[sort&compress]{natbib}
\usepackage{authblk}
\usepackage[misc]{ifsym}

\def\d{\mathrm d}

\def\const{\operatorname {const}}
\def\sign{\operatorname {sign}}

\def\arccosh{\operatorname{arccosh}}

\def\sign{\operatorname{sign}}
\def\const{\operatorname{const}}

\def\Res{\operatorname{Res}}

\theoremstyle{remark}
\newtheorem{remark}{\it Remark}

\def\Re{\operatorname{Re}}
\def\Im{\operatorname{Im}}
\def\I{\mathrm i}

\def\cc{\mathrm{c.c.}}

\def\Idva{(\mathcal I_n^N)^\mathrm{pass}}
\def\Isss{(\mathcal I_n^N)^\mathrm{stop}}
\def\Idvaa{(I_{n-N})^\mathrm{pass}}
\def\Issss{(I_{n-N})^\mathrm{stop}}
\def\VNNPass{(\VF_n^N)^\mathrm{pass}}
\def\bVNNPass{(\breve{\VF}_n^N)^\mathrm{pass}}
\def\bVNNStop{(\breve{\VF}_n^N)^\mathrm{stop}}
\renewcommand{\=}{\stackrel{\mbox{\scriptsize def}}{=}}
\def\asst{{\mathrm s}}
\def\pFO{}

\def\EXP#1{\mathrm e^{#1}}
\def\UF{{\mathscr U}}
\def\VF{{\mathscr V}}
\def\VFF{\breve{\mathscr V}}
\def\TT{\mathscr T}

\let\mathscr=\mathcal

\def\de{\delta}

\def\DM{m-1}
\def\E{E_n^N(\Omega)}
\def\Es{E_n^N(\Omega)}
\def\k{{N-n}}
\def\FFF{F}

\makeatletter

\@addtoreset{equation}{section}
\@addtoreset{remark}{section}
\makeatother

\begin{document}
\selectlanguage{english}

\title{Non-stationary elastic wave scattering and energy transport 
in a one-dimensional harmonic chain with an isotopic defect}


\author{Serge N. Gavrilov}
\author{Ekaterina V. Shishkina}
\affil{Institute for Problems in Mechanical Engineering RAS,\\ St.~Petersburg, Russia}

%


\maketitle

\begin{abstract}
The fundamental solution describing non-stationary
elastic wave scattering on an isotopic defect 
in a one-dimensional harmonic chain is obtained in an asymptotic form. 
The chain is subjected to unit impulse point loading applied to a particle far
enough from the defect.  The solution is a large time asymptotics at a moving
point of observation, and it is in excellent agreement with the corresponding numerical
calculations. At the next step, we assume that the applied point impulse excitation 
has random amplitude.  This allows one to model the heat transport 
in the chain and across the defect as the transport of the mathematical
expectation for the kinetic energy and to use
the conception of the kinetic temperature. 
To provide a simplified continuum description for this process,
we separate the slow in time component
of the kinetic temperature. This quantity can be calculated using the asymptotics
of the fundamental solution for the
deterministic problem.  We
demonstrate that there is a thermal shadow behind the defect: the order of
vanishing for the slow temperature is larger for the particles behind the
defect than for the particles between the loading and the defect.
The presence of the thermal shadow is related to a non-stationary wave phenomenon,
which we call the anti-localization of non-stationary waves.
Due to the presence of the shadow, the continuum slow kinetic temperature 
has a jump discontinuity at the defect. Thus, the system under
consideration can be a simple model for the non-stationary phenomenon,
analogous to one characterized by the Kapitza thermal resistance. 
Finally, we analytically
calculate the non-stationary transmission function, which describes the
distortion (caused by the defect) of the slow kinetic temperature profile at a far zone
behind the defect.
\end{abstract}


\section{Introduction}
We consider the non-stationary dynamics of a one-dimensional
harmonic crystal (a linear chain) with an isotopic defect subjected to a
point impulse loading applied outside (more precisely, far enough) of the defect. 
The present paper is a
natural continuation of our previous paper \cite{Shishkina2023cmat}, where we
considered the
same system but with the loading applied at the defect. Now, the alternated 
mathematical formulation corresponds to the essentially different physical nature
of the dynamical
processes in the chain. Namely, we deal with the problem of non-stationary 
scattering (or non-stationary diffraction) of an elastic quasi-wave on a point
defect. We speak
about a quasi-wave since the perturbations in discrete systems propagate at
infinite speed. In what follows, we often refer
to quasi-waves as waves. On the other hand, the problem concerning non-stationary scattering is 
essentially more difficult from a mathematical point of view compared with
the one
considered in \cite{Shishkina2023cmat}. The practical motivation for our
study is experimental observations (see, e.g., \cite{Chen2012}), which
indicate
that the presence of isotopic defects essentially affects the thermal conductivity
of pure nanomaterials.

An extensive bibliography on the studies, which deal with a chain with an
isotopic defect, can be found in our previous paper \cite{Shishkina2023cmat}.
Many of them consider the case when the loading (deterministic or stochastic)
is applied at the defect
\cite{Teramoto1960,Kashiwamura1962,hemmer1959dynamic,magalinskii1959dynamical,Mueller1962,Mueller2012,Turner1960,Rubin1960,Rubin1961,Rubin_1963,Lee1989,Yu2019}. In
\cite{Takizawa1967,Takizawa1968}, the general solution is obtained, which
involves both cases when the loading is applied on or outside the defect, but
it has a complicated form and is difficult for analysis.
In 
\cite{kannan2013heat,Paul2020,Gendelman2021,Plyukhin2020}, the 
steady-state problem concerning the thermal conductivity of a
chain of a finite length with an isotopic defect subjected to thermal
(stochastic) sources at the ends is considered. The scattering problem was
considered mostly in the stationary statement
\cite{Koster1954a,Lifsic1956,Fellay1997,Kosevich2008,Kosevich1997,Kossevich1999,Lifshitz1966}.
In particular, the stationary three-dimensional problems concerning the scattering on a
point or a two-dimensional (plane) defect are discussed in 
\cite{Kossevich1999,Lifshitz1966}. In
\cite{Fellay1997,Kosevich2008,Kosevich1997},
one-dimensional problems
concerning scattering in a chain with a defect of a complex structure, where
multichannel propagation is possible, are considered in the stationary
statement. The similar, from a mathematical point of view, problem  concerning a chain with an interface
is discussed in the stationary formulation in 
\cite{Jex1986,Kakodkar2015,Kuzkin2023,Lumpkin1978,Polanco2013,Saltonstall2013,Steinbruechel1976}.
Note that in physics, the phenomenon discussed in the framework of stationary
problems is often referred to as phonon scattering 
\cite{Kakodkar2015,Kosevich2008,Steinbruechel1976,Saito2018}. The incident wave is
generally taken as an infinite harmonic plane wave, i.e., a phonon, not a
harmonic wave corresponding to a point harmonic load. The stationary solution for
a semi-infinite chain harmonically excited at the free end with an isotopic defect at an
arbitrary position is obtained in \cite{Mokole1990}.

In the current paper, as well as in \cite{Shishkina2023cmat}, we consider two
problems: a deterministic and a stochastic one. 
We initially formulate
the scattering problem as a deterministic problem 
(Sect.~\ref{sec-formulation1}). Since a unit impulse point load
is under consideration, the obtained solution can be treated 
as the fundamental one. Since the main point of interest in the paper is
related to the stochastic kinetic energy (heat)
transport problem, we are interested in the derivation of 
the fundamental solution for the particle velocities. 
In the framework of the stochastic problem (Sect.~\ref{sec-formulation2}), we
assume that the applied point impulse excitation has a random amplitude. This allows one
to model the heat transport in the chain and across the defect as the transport 
of the mathematical expectation of the kinetic energy 
and to use
the conception of the kinetic temperature. The propagation of the kinetic
temperature is referred to in the paper as the thermal motions. The
non-stationary fundamental
solution, which describes this process, is the thermal fundamental solution.

Due to the linearity of the
problem, the solution of the
scattering problem can be represented as a superposition of an incident wave,
which corresponds to the same loading in the uniform chain,
and a scattered wave. The corresponding 
decomposition for the Green functions in the frequency domain is obtained in
Sect.~\ref{App-with}.
In Sect.~\ref{sect-scattering},
we obtain the asymptotic solutions, which describe the non-stationary
scattered wave and the total wave-field.
The obtained solutions provide a continuum description of the wave-field and
are in excellent agreement with the corresponding numerical calculations.
The non-stationary solution for the scattered
wave-field is found as a large time asymptotics at
a moving point of observation, using a technique analogous to one used in
our previous studies
\cite{Shishkina2023jsv,Shishkina2023cmat,Gavrilov2022ijhmt}. 

In Sect.~\ref{sect-transport}, we start to deal with the energy (heat) transport
problem.
Following the procedure suggested
in \cite{Gavrilov2022ijhmt}, we use the asymptotics
of the fundamental solution for the
deterministic problem and get
slow and fast decoupling of the thermal fundamental solution.
The slow motion (the slow component of the kinetic temperature) is introduced as
the time average of the kinetic temperature over fast phases (time-like
variables). Note that
initially, the slow motion was introduced
\cite{krivtsov2015heat,Kuzkin2017fast} as a formal
solution of equations for covariances with dropped high-order
time derivatives. 
The slow motion
provides a simplified continuum 
description for the heat transport process. The fast motion is the energy
oscillation related to the transformation of kinetic energy into 
potential energy and back \cite{Gavrilov2019PhysRevE,krivtsov2014energy,Kuzkin2017fast}. 
In discrete harmonic systems,
where a stochastic loading is distributed in space
\cite{krivtsov2015heat,krivtsov-da70,Sokolov2021,Kuzkin2017fast,Kuzkin2019} or in time
\cite{Gavrilov2019cmat,Gavrilov2022cmat,Gavrilov2020cmat}, according to
numerical calculations,
the fast motion vanishes. Thus, we expect that to describe heat
transport for such a loading, it is enough to calculate the 
convolution of the loading with the 
fundamental solution 
 for the slow motion
obtained in the paper 
(without any
spatial or temporal averaging).

In Sect.~\ref{sec-anti}, we
demonstrate that there is a thermal shadow behind the defect: the
order of
vanishing for the slow temperature is larger for the particles behind the
defect than for the particles between the loading and the defect.
The presence of the thermal shadow is related to a non-stationary wave phenomenon,
which we call the anti-localization of non-stationary waves 
\cite{Shishkina2023cmat,Shishkina2023jsv,Gavrilov2023DD}.
Due to the presence of the shadow, the continuum slow kinetic temperature 
has a jump discontinuity at the defect. Thus, we have shown that the system under
consideration can be a simple model for the non-stationary phenomenon
analogous to one characterized by the Kapitza (interfacial) thermal resistance
\cite{Gendelman2021,Kapitza1941,Lumpkin1978,Paul2020}. Finally, we analytically
calculate the non-stationary transmission function, which describes the
distortion (caused by the defect) of the slow kinetic temperature profile at a far zone
behind the defect.

Summarizing, the novelty of our paper is related to the non-stationary
formulation of the problems under consideration and to the solutions obtained
in asymptotic form, which describe the non-stationary
effects, namely, the thermal shadow behind the defect and the phenomenon of
the temperature jump analogous 
to one characterized by the Kapitza thermal resistance.


\section{Nomenclature}
\label{sec-notation}
In the paper, we use the following general notation:
\begin{description}	
\item[$\mathbb Z$] is
the set of all integers;
\item[$\mathbb R$] is
the set of all real numbers; 
\item[$H(\cdot)$] is the Heaviside step-function;
\item[$\langle\cdot\rangle$] is the mathematical expectation for a random quantity;
\item[$\de_n$ ] is {the Kronecker delta} ($1$ if and only if $n=0$, $0$ otherwise, {$n\in\mathbb Z$});
\item[$\de(t)$ ] is {the Dirac delta-function}; 
\item[$k_B$] is the dimensionless Boltzmann constant (without loss of
generality one can take $k_B=1$);
\item[$J_n(\cdot)$] is the Bessel function of the first kind of order $n$;
\item[$\Gamma(\cdot)$] is the Gamma function;
\item[$\cc$] are the complex conjugate terms;
\item[$m$] is the dimensionless mass of an isotopic defect;
\item[$n$] is a particle number;
\item[$N$] is the number of the particle, where the loading is applied.

\end{description}	

\section{The problem formulation}
\label{sec-formulation}
\subsection{Non-stationary elastic wave scattering}
\label{sec-formulation1}
Consider a chain of point particles of an equal mass with one alternated mass. 
All masses are connected by linear springs with the same stiffness.
The equations of motion in the dimensionless form can be expressed as the following 
infinite system of differential-difference equations:
\begin{gather}
m_n\ddot{u}_n-(u_{n+1}-2u_n+u_{n-1})= \delta_{n-N}\, p(t).
\label{chain-eq-basic-al0}
\end{gather}
Here 
$n \in \mathbb{Z}$,  
$u_n(t)$ is the dimensionless displacement of the particle with a number
$n$,
$m_n$ is
the dimensionless mass of a particle with a number $n$:
\begin{gather}
m_n=1+\delta_n(m-1),
\label{m_n}
\end{gather}
overdot denotes the derivative with respect to
the dimensionless time $t$. We assume that the dimensionless mass of the defect
particle $m=m_0$ is such that
\begin{gather}
m>0, \qquad m\neq1.
\end{gather}
The external force $p(t)$ is applied to the 
mass point with the number $N=\const$, $N\neq0$. 
Without loss of generality, in what follows, we assume that $N>0$.

Taking into account Eq.~\eqref{m_n},
Eq.~\eqref{chain-eq-basic-al0} 
can be rewritten in the following form:
\begin{gather}
\ddot{u}_n-(u_{n+1}-2u_{n}+u_{n-1})=\delta_{n-N}\, p(t) - \delta_n\,
(\DM) \ddot{u}_0.
\label{chain-gov-eq-loaded}
\end{gather}
The differential-difference operator in the left-hand side of 
Eq.~\eqref{chain-gov-eq-loaded} corresponds to a 
uniform chain of mass points of unit mass connected by springs of unit
stiffness.

\begin{remark}  
In the paper, we use the dimensionless problem formulation from the very beginning.
Non-dimensionalization is discussed, e.g., in \cite{Shishkina2023cmat}.
\end{remark}

Since we are interested mostly in thermal processes, which are related to
the propagation of the kinetic energy (or kinetic temperature), 
in what follows, we deal with the
expression for the particle velocity $\dot u_n$.
In the paper, we use the fundamental solution 
\begin{equation}
u_n=\UF_n^N,
\qquad
\dot u_n=\VF_n^N\=\dot\UF_n^N
\end{equation}
of the
deterministic problem, which corresponds to the choice of the external force 
as the pulse force
\begin{equation}
p(t)=\delta(t).
\label{p-detlta-t}
\end{equation}
In this case, the initial conditions for Eq.~\eqref{chain-eq-basic-al0}
can be
formulated in the following form, which is conventional for distributions (or
generalized functions) \cite{Vladimirov1971}:
\begin{equation}
u_n \big|_{t<0} \equiv 0.
\label{initial-cond}
\end{equation}

The generalized initial value problem 
\eqref{chain-eq-basic-al0} with the right-hand side 
defined by Eq.~\eqref{p-detlta-t}
and initial conditions \eqref{initial-cond}
can be equivalently formulated in the form of a classical initial value problem
for the system of equations
\begin{equation}
m_n\ddot{u}_n-(u_{n+1}-2u_n+u_{n-1})= 0
\label{chain-eq-basic-class}
\end{equation}
with initial conditions in the classical form \cite{Vladimirov1971}
\begin{gather}
u_n(0)=0,\qquad \dot u_n(0)=\delta_{n-N}.
\end{gather}

In the particular case $m=1$ of a uniform chain, the exact expression for the fundamental solution 
$\VF_n^N\big|_{m=1}$ is $V_{n-N}$ \cite{schrodinger1914dynamik}, where 
\begin{equation}
V_{n}=J_{2n}(2t)=J_{2|n|}(2t), \quad n\in\mathbb Z.
\label{Sro-bessel}
\end{equation}
The fundamental solution 
$\mathscr V_n\= \mathscr V_n^0$
of problem
\eqref{chain-eq-basic-al0}, \eqref{initial-cond}
wherein $N=0$, and the external force is defined by
\eqref{p-detlta-t}
was asymptotically investigated in our previous work 
\cite{Shishkina2023cmat}.

\subsection{Kinetic energy (heat) transport}
\label{sec-formulation2}
Consider the case of a point random initial excitation.
{Let the initial conditions for Eq.~\eqref{chain-eq-basic-class} be as
follows:}
\begin{gather}  
u_n(0)=0,\qquad \dot u_n(0)=\rho{\delta_{n-N}}.
\end{gather}
Here $n\in\mathbb Z$, $\rho$ is a random quantity such that 
\begin{gather}
\langle\rho\rangle=0,\qquad \langle\rho^2\rangle=\sigma.
\label{sol-cov-general}
\end{gather}
The (dimensionless) kinetic temperature $\varTheta_n$ is conventionally introduced by the
following formula:
\begin{equation}
\varTheta_n\=
2k_B^{-1}\langle K_n\rangle,
\label{sol-T-def}
\end{equation}
where 
\begin{gather}
K_n(t)=\frac{m_n\dot u_n^2(t)}2
\end{gather}
is the kinetic energy,
\begin{gather}
\langle K_n(t)\rangle=
\frac{m_n}2\,\langle \dot u^2_n \rangle 
=
\frac{\sigma m_n}2\, \big(\VF_n^N(t)\big)^2
\end{gather}
is the mathematical expectation for the kinetic energy,
\begin{gather}
\mathscr E
\=
\sum_n\langle K_n(0)\rangle=
\frac{\sigma }2\, \big(\VF_0^N(0)\big)^2
=
\frac{\sigma }2\, 
\label{E-def}
\end{gather}
is the mathematical expectation for the initial kinetic (as well as total)
energy for the whole harmonic crystal.
Thus,
\begin{equation}
\langle K_n(t)\rangle=\mathscr E m_n  \big(\VF_n^N(t)\big)^2,
\end{equation}
and, therefore,
\begin{gather}
\varTheta_n(t)=
k_B^{-1} \mathscr E 
\,\TT_n^N(t),
\label{Tc-def}
\end{gather}
where we call the quantity
\begin{gather}
\TT_n^N(t)=
2 m_n \big(\VF_n^N(t)\big)^2
\label{thermal-fs}
\end{gather}
the thermal fundamental solution. One has 
\begin{equation}
\TT_n^N(0)=2\delta_{n-N}.
\label{2dn}
\end{equation}
We choose the fundamental solution $\TT_n^N$, which satisfies the initial
normalization condition \eqref{2dn}, to be in agreement with the previous studies 
\cite{krivtsov2015heat,krivtsov-da70,Gavrilov2019cmat,Sokolov2021,Gavrilov2022ijhmt},
where a uniform chain is under consideration. 
In that particular case $m=1$, the exact expression for 
the thermal fundamental solution $\TT_n^N\big|_{m=1}=T_{n-N}$
\cite{hemmer1959dynamic,Sokolov2021,Gavrilov2022ijhmt}, where 
\begin{equation}
T_{n}\=2J^2_{2n}(2t)=2J^2_{2|n|}(2t),\quad n\in \mathbb Z.
\label{TnN-def}
\end{equation}
The thermal fundamental solution 
$\mathscr T_n\= \mathscr T_n^0$
for problem
\eqref{chain-eq-basic-al0}, \eqref{initial-cond}
wherein $N=0$, and the external force is defined by
\eqref{p-detlta-t},
was asymptotically investigated in our previous work 
\cite{Shishkina2023cmat}.


\color{black}
\section{The Green function in the frequency domain}
\label{App-with}
Consider now Eq.~\eqref{chain-gov-eq-loaded}, 
wherein 
\begin{gather}
u_n(t) = U_n(\Omega)\, \mathrm{e}^{-\I \Omega t},
\label{u-harmonic}
\\
p(t)=\mathrm e^{-\I\Omega t}.
\label{harmonic-Uscr}
\end{gather}
We substitute expressions~\eqref{u-harmonic}, \eqref{harmonic-Uscr}
into Eq.~\eqref{chain-gov-eq-loaded}.
This yields
\begin{equation}
(-\Omega^2+2)U_n-U_{n+1}-U_{n-1}=\delta_{n}\Omega^2(\DM)U_0+\delta_{n-N}.
\label{chain-gov-eq-loaded-Four1}
\end{equation}
The boundary conditions are assumed to be 
in the form of the Sommerfeld radiation conditions at
infinity for $\Omega\in\mathbb P$ and
the vanishing boundary conditions for $\Omega\in\mathbb S$. Here 
$
\mathbb P\=[-\Omega_\ast,\Omega_\ast] 
$
is the pass-band,
$
\mathbb S\=(-\infty,-\Omega_\ast)\cup(\Omega_\ast,\infty)
$ 
is the stop-band,
$
\Omega_\ast\=2
$
is the cut-off (or boundary) frequency.
The corresponding steady-state solution $U_n=\mathscr G_{n}^N$ of the obtained
equation is the Green function in the frequency domain for a chain with an
isotopic defect at $n=0$ and load at $n=N$.

Since Eq.~\eqref{chain-gov-eq-loaded-Four1} is a linear difference equation,
the solution 
$U_n=\mathscr G_n^N$
can be represented as the superposition of the incident
harmonic quasi-wave $G_{n-N}$ and the scattered one $\breve{\mathscr G}_n^N$:
\begin{equation}
\mathscr G_n^N=G_{n-N}+\breve{\mathscr G}_n^N.
\label{Green-sum}
\end{equation}
The components in the right-hand side of 
Eq.~\eqref{Green-sum} are individual reactions of the system described by 
Eq.~\eqref{chain-gov-eq-loaded-Four1} on the two components of loading in the
right-hand side of 
Eq.~\eqref{chain-gov-eq-loaded-Four1}.
The first component $G_{n-N}$, which corresponds to 
$\delta_{n-N}$,
is defined by \cite{Shishkina2023cmat} 
\begin{equation}
G_n(\Omega)\=\mathscr G_n^0(\Omega)\big|_{m=1},
\label{G-def}
\end{equation}
where
\begin{align}
&
\mathscr G_n^0(\Omega)=\frac{\EXP{\I a(\Omega) |n|\sign \Omega}}
{-m\Omega^2-2\EXP{\I a(\Omega) \sign \Omega}+2}
,&\quad
&\Omega\in\mathbb P;
\label{Green-function0-lower}
\\
&
\mathscr G_n^0(\Omega)=
\frac{(-1)^{|n|}\EXP{-b(\Omega)|n|}}{-m\Omega^2+2\EXP{-b(\Omega)}+2}
,&\quad
&\Omega\in\mathbb S
\label{Green-function0-upper}
\end{align}
is the corresponding solution of
Eq.~\eqref{chain-gov-eq-loaded-Four1} in the case $N=0$.
Here 
\begin{gather}  
a(\Omega)=\arccos \frac{2-\Omega^2}2,
\label{wn-pass}
\end{gather}
is the absolute value of the wave-number $q=\pm a(\Omega)$ in the pass-band,
\begin{gather}  
b(\Omega)=
\arccosh\frac{\Omega^2-2}{2}
\label{wn-stop}
\end{gather}
is the absolute value for the imaginary
part of the wave-number $q(\Omega)=\pi\pm\I b(\Omega)$ in the stop-band.
The frequency $\Omega$ and the wave-number $q$
satisfy the dispersion relation for a uniform chain
\cite{Shishkina2023cmat,Montroll1955}
\begin{equation}
\Omega^2=4\sin^2\frac{q}{2}\equiv 2(1-\cos q).
\label{dis-form3-al0}
\end{equation}
Thus, the incident quasi-wave $G_{n-N}$ is the Fourier transform with respect to
time of the fundamental solution
$V_{n-N}$ defined by Eq.~\eqref{Sro-bessel}, whereas 
$\mathscr G_n^N$ is the Fourier transform of $\mathscr V_n^N$.

The scattered quasi-wave $U_n=\breve{\mathscr G}_n^N$, which corresponds to
the term $\delta_{n}\Omega^2(\DM)U_0$ in the right-hand side of 
Eq.~\eqref{chain-gov-eq-loaded-Four1} 
now can be expressed as
\begin{equation}
\breve{\mathscr G}_n^N=\Omega^2(m-1)G_N\mathscr G_n^0.
\label{Green-breve}
\end{equation}

Substituting Eq.~\eqref{wn-pass} and Eq.~\eqref{wn-stop}
into Eq.~\eqref{Green-function0-lower}
and Eq.~\eqref{Green-function0-upper}, respectively, one can equivalently
rewrite $\mathscr G_n$ in the following form \cite{Shishkina2023cmat}:
\begin{align}
&
\mathscr G_n^0(\Omega)=-\frac{
\EXP{\I  |n|\sign \Omega\, \arccos\frac{2-\Omega^2}2}}
{(\DM)\Omega^2+\I\Omega \sqrt{4-\Omega^2}}
,&\quad
&\Omega\in\mathbb P;
\label{Green-function0-lower-e}
\\
&
\mathscr G_n^0(\Omega)=\frac{(-1)^{|n|}2^{|n|}}
{\Phi^{|n|-1}(\Omega)\big((-m\Omega^2+2)\Phi(\Omega)+4\big)}
,&\quad
&\Omega\in\mathbb S;
\label{Green-function0-upper-e}
\end{align}
where
\begin{equation}
\Phi(\Omega)\=\Omega^2-2+|\Omega|\sqrt{\Omega^2-4}.
\end{equation}

\section{Non-stationary scattering}
\label{sect-scattering}
The fundamental solution for the particle velocity $\VF_n^N$ can be represented
as the inverse Fourier transform:
\begin{multline}
\VF_n^N
=
(\VF_n^N)^\mathrm{pass}
+
(\VF_n^N)^\mathrm{stop}
\=
-\frac{\I}{2\pi}\left(\int_{\mathbb P}+\int_{\mathbb S}\right)
\Omega
{\mathscr G}_n^N
\EXP{-\I\Omega t}
\d \Omega
\\
=
-\frac{\I}{2\pi}\left(\int_{\mathbb P_+}+\int_{\mathbb S_+}\right)
\Omega
{\mathscr G}_n^N
\pFO\EXP{-\I\Omega t} \, \d\Omega+\cc
\label{V-1whole}
\end{multline}
where $\mathscr G_n^N$ is defined by 
Eqs.~\eqref{Green-sum}--\eqref{Green-breve},
\begin{equation}
\mathbb P_+\=[0,\Omega_\ast], 
\qquad
\mathbb S_+\=(\Omega_\ast,\infty).
\end{equation}
Since, by construction, $-\I\Omega G_{n-N}$ is the Fourier transform for $V_{n-N}$ 
defined by \eqref{Sro-bessel}, one has
\begin{multline}
V_{n-N}=
(V_{n-N})^\mathrm{pass}+(V_{n-N})^\mathrm{stop}\=
-\frac{\I}{2\pi}\left(\int_{\mathbb P}+\int_{\mathbb S}\right) 
\Omega G_{n-N}\,\EXP{-\I\Omega t}\, \d\Omega
\\=
-\frac{\I}{2\pi}\left(\int_{\mathbb P_+}+\int_{\mathbb S_+}\right) 
\Omega G_{n-N}\,\EXP{-\I\Omega t}\, \d\Omega+\cc
=
\Idvaa+\Issss+\cc 
\label{Vn-pass-stop}
\end{multline}
Thus,
\begin{equation}
\VF_n^N=V_{n-N}+\breve \VF_n^N
\label{VF-super}
\end{equation}
where
\begin{multline}
\breve\VF_n^N
=
(\breve\VF_n^N)^\mathrm{pass}
+
(\breve \VF_n^N)^\mathrm{stop}
\=
-\frac{\I}{2\pi}\left(\int_{\mathbb P}+\int_{\mathbb S}\right)
\Omega
\breve{\mathscr G}_n^N
\EXP{-\I\Omega t}
\d \Omega
\\
=
-\frac{\I}{2\pi}\left(\int_{\mathbb P_+}+\int_{\mathbb S_+}\right)
\Omega
\breve{\mathscr G}_n^N
\pFO\EXP{-\I\Omega t} \, \d\Omega+\cc
=
\Idva+\Isss+\cc 
,
\label{bVNNPass-Stop-def}
\end{multline}
where $\breve{\mathscr G}_n^N$ is defined by 
\eqref{Green-breve}.

To estimate 
the integrals 
$\Idva$ and $\Isss$, we use, in what follows, the procedure of asymptotic
evaluation for large times based on the method of stationary
phase \cite{erdelyi1956asymptotic,Fedoryuk1977,temme2014}. 

%

\subsection{The contribution from the pass-band}
The incident wave-field $V_{n-N}$ defined by 
Eq.~\eqref{Sro-bessel}
is an even function of the discrete variable
$n-N$ \cite{Gavrilov2022ijhmt}. To have the possibility to estimate analytically the
total wave-field
\eqref{VF-super} superposed of the incident and scattered wave-fields,
we use, in what follows, the asymptotic representation of $V_{n-N}$ on the
moving at a speed $0\leq w<1$ observation point 
\begin{equation}
|n-N|=wt.
\label{moving-front}
\end{equation}
It has the following form \cite{Gavrilov2022ijhmt,Shishkina2023cmat}:
\begin{equation}
V_{n-N}
=\frac{H(1-w)}{\sqrt{\pi t}\sqrt[4]{1-w^2}}
\cos\Big(
\omega
t+\frac{\pi}4\Big)
+O(t^{-1}),
\label{ins-wave-asymp11}
\end{equation}
where
\begin{gather}
\omega \=2(w\arccos w -\sqrt{1-w^2}).
\label{phi_s}
\end{gather}
After the backward substitution $w=|n-N|/t$
\cite{Gavrilov2022ijhmt,Shishkina2023cmat},
Eq.~\eqref{ins-wave-asymp11} can be rewritten as
\begin{equation}
V_{n-N}
\\=\frac{H_-
\cos\left(\omega_-t+\frac\pi4\right)
}{\sqrt\pi\sqrt[4]{t^2-(n-N)^2}}
+O(t^{-1}),
\label{ins-wave-asymp22}
\end{equation}
where
\begin{gather}
\omega_\pm\=\omega\Big|_{w={\frac{|n\pm N|}t}}
={2\bigg(\frac{|n\pm N|}t\arccos\frac{|n \pm N|}t 
-\frac{\sqrt{t^2-(n\pm N)^2}}t\bigg)},
\label{phi_s-pm}
\\
H_\pm\=H\big(t-|n\pm N|\big).
\label{H-pm}
\end{gather}
At the same time, the scattered component $\VFF_n^N$
is an even function of the variable $n$.

Now we plan to investigate the behaviour of $\bVNNPass$ and
\begin{equation}
\VNNPass=V_{n-N}+\bVNNPass
\label{V-total}
\end{equation}
inside two intervals.
The first one is $n\leq0$, where the
scattered wave $\bVNNPass$ is a left-running wave,
the second one is $n>0$, where $\bVNNPass$ is a right-running wave.

%

\subsubsection{$n\leq0$}
\label{sect-n<0}
First, we should treat $n\leq0$. One has:
\begin{equation}
\Idva=-\frac{\I}{2\pi}\int_{0}^{2} 
\frac{(m-1)\Omega\,{\mathrm e}^{\I(|n|+N)\arccos\frac{2-\Omega^2}2-\I\Omega t}}
{-(4-\Omega^2)+\I(m-1)\Omega\sqrt{4-\Omega^2}} \, \d\Omega,
\label{I-pass-left-pre}
\end{equation}
In this case, we plan to estimate the integral in the right-hand side of 
\eqref{I-pass-left-pre} at the same moving fronts 
\eqref{moving-front} as we have done for $V_{n-N}$ in
Eq.~\eqref{ins-wave-asymp11}:
\begin{equation}
|n-N|=|n|+N=wt.
\label{wt_for-n<0}
\end{equation}
\begin{remark}	
The choice of the moving observation fronts is generally ambiguous.
One can try to
construct asymptotics on moving observation points different from 
Eq.~\eqref{wt_for-n<0} (see Appendix~\ref{app-C}). However, those asymptotics seem 
practically inapplicable as the approximate solution in terms of variables $n$
and $t$.
\end{remark}
We represent integral 
\eqref{I-pass-left-pre}
as follows:
\begin{equation}
\Idva=-\frac{\I }{2\pi}\int_{0}^{2}A^{\mathrm{pass}}(\Omega)\EXP{\phi(\Omega)t} \, \d
\Omega,
\label{I-pass-left}
\end{equation}
where
\begin{gather}
A^{\mathrm{pass}}(\Omega)=\frac{(m-1)\Omega }{-(4-\Omega^2)+\I(m-1)\Omega\sqrt{4-\Omega^2}},
\label{A-pass-upright-def}
\\
\phi(\Omega)=w\arccos\frac{2-\Omega^2}2-\Omega,
\label{phase-phi(Omega)}
\end{gather}
are the amplitude and the phase, respectively.
Applying the stationary phase method
analogously to \cite{Gavrilov2022ijhmt,Shishkina2023cmat},
one can derive the following asymptotics:
\begin{equation}
\Idva=-\frac{\I 
A_{\mathrm s}\EXP{\I(\omega t+\frac{\pi}4)}
}{\sqrt{2\pi|\phi''(\Omega_\asst)
|t}}+O(t^{-1}),
\label{int-as}
\end{equation}
where
\begin{equation}
\Omega_{\mathrm s}=2\sqrt{1-w^2}
\label{st-point}
\end{equation}
is the unique non-degenerate stationary point, which exists for $0<w<1$;
\begin{equation}
\phi''(\Omega_\asst)=\frac{\sqrt{1-w^2}}{2w^2};
\end{equation}
$\omega = \phi(\Omega_s)$ is defined by Eq.~\eqref{phi_s};
\begin{equation}
A_{\mathrm s} \= A^{\mathrm{pass}}(\Omega_\asst)=
\frac{(m-1)\sqrt{1-w^2}}{2w(-w+\I(m-1)\sqrt{1-w^2})}.
\end{equation}
Now, one has
\begin{multline}
(\VFF_n^N)^{\mathrm{pass}}=\Idva+\cc=-\frac{\I H(1-w) }{\sqrt{2\pi|\phi''(\Omega_\asst)
|t}}\left(A_{\mathrm s}\EXP{\I(\omega t+\frac{\pi}4)}-\Bar{A}_{\mathrm s}\EXP{-\I(\omega
t+\frac{\pi}4)}   \right) +O(t^{-1})
\\
=\frac{2 H(1-w)}{\sqrt{2\pi|\phi''(\Omega_\asst)
|t}}\left( \Im A_{\mathrm s} \cos\Big(\omega t+\frac{\pi}4\Big )+\Re A_{\mathrm s} \sin\Big
(\omega t+\frac{\pi}4 \Big )  \right)
+O(t^{-1}).
\end{multline}
The Heaviside step-function in the last equation
indicates that there are no stationary points for $w>1$.
Calculating $\Re A_{\mathrm s}$, $\Im A_{\mathrm s}$, one gets:

\begin{equation}
\Re A_s =\frac{-(m-1)w\sqrt{1-w^2}}{2w(w^2+(m-1)^2(1-w^2))},
\end{equation}
\begin{equation}
\Im A_s =\frac{-(m-1)^2(1-w^2)}{2w(w^2+(m-1)^2(1-w^2))}.
\end{equation}
Now, one obtains:
\begin{multline}
\bVNNPass=-\frac{H(1-w)(m-1)\sqrt[4]{1-w^2}}{\sqrt{\pi t}(w^2+(m-1)^2(1-w^2))}
\Big((m-1)\sqrt{1-w^2}\cos\Big(\omega t+\frac{\pi}4\Big )
\\
+w\sin\Big (\omega t+\frac{\pi}4 \Big ) 
\Big)
+O(t^{-1}).
\label{asymp-J_2}
\end{multline}
The last expression can be transformed into the following form:
\begin{equation}
\bVNNPass
=-\frac{H(1-w)(m-1)\sqrt[4]{1-w^2}}{\sqrt{\pi t}\sqrt{w^2+(m-1)^2(1-w^2)}}
\sin\Big(\omega t+\psi+\frac{\pi}4
\Big)
+O(t^{-1})
,
\label{asymp-J_2-mod}
\end{equation}
where
\begin{equation}
\psi=\arctan\frac{(m-1)\sqrt{1-w^2}}{w}.
\label{psi}
\end{equation}
Now, following \cite{Gavrilov2022ijhmt,Shishkina2023cmat}, we obtain the
approximate solution by the backward substitution $w=|n|/t$:
\begin{equation}
\bVNNPass
=-\frac{H_-(m-1)\sqrt[4]{t^2-(n-N)^2}}{\sqrt{\pi }
\sqrt{(n-N)^2+(m-1)^2(t^2-(n-N)^2)}}
\sin\Big(\omega_- t+\psi_-+\frac{\pi}4
\Big)
+O(t^{-1}) ,
\label{bVNNPass1}
\end{equation}
where
\begin{gather}
\psi_\pm\=\psi\Big|_{w=\frac{|n\pm N|}t}=
\arctan\frac{(m-1)\sqrt{t^2-(n\pm N)^2} }{|n\pm N|}.
\label{psi-n<0}
\end{gather}

To obtain the expression for the total transmissive wave-field $\VNNPass$,
now we substitute 
\eqref{ins-wave-asymp11}
and 
\eqref{asymp-J_2}
into the right-hand side of Eq.~\eqref{VF-super}.
This yields:
\begin{multline}
(\VF_n^N)^\mathrm{pass}=
\frac{H(1-w)w}{\sqrt{\pi t}\sqrt[4]{1-w^2}\sqrt{w^2+(m-1)^2(1-w^2)}}
\cos\Big(\omega t+\psi+\frac{\pi}4 \Big)+O(t^{-1})
\\
=
\frac{H_-|n-N| \cos\big(\omega_- t+\psi_-+\frac{\pi}4 \big) }
{
\sqrt{\pi}\sqrt[4]{t^2-(n-N)^2}
\sqrt{(n-N)^2+(m-1)^2(t^2-(n-N)^2)}
}
+O(t^{-1}).
\label{dotu-pass-n<0}
\end{multline}


\subsubsection{$n>0$}
\label{difn>0}
Consider now the case $n>0$. 
The scattered wave-field $\bVNNPass$ must be an even function of $n$. Thus, to
obtain the expression for $\bVNNPass$ we now need to substitute $-n$ instead of
$n$ into Eq.~\eqref{bVNNPass1}. This yields
%
%
\begin{equation}
\VFF_n^N=
-\frac{H_+(m-1)\sqrt[4]{t^2-(n+N)^2}}{\sqrt{\pi
}\sqrt{(n+N)^2+(m-1)^2(t^2-(n+N)^2)}}\sin\Big(\omega_+ t+\psi_++\frac{\pi}4
\Big)
+O(t^{-1}).
\label{asymp-J_2-2}
\end{equation}
where $\omega_+,\ H_+,\ \psi_+$ are defined by Eqs.~\eqref{phi_s-pm}, \eqref{H-pm},
\eqref{psi-n<0}, respectively.

The expression for total wave-field $\VNNPass$ now can be calculated by Eq.~\eqref{V-total},
wherein $\bVNNPass$ and $V_{n-N}$ are given by 
Eqs.~\eqref{asymp-J_2-2} and 
\eqref{Sro-bessel}, respectively. Using approximate solution
\eqref{ins-wave-asymp22} instead of exact one~\eqref{Sro-bessel} results in
\begin{multline}
\VNNPass\simeq
\frac{H_-}{\sqrt{\pi}\sqrt[4]{t^2-(n-N)^2}}
\cos\Big(\omega_- t+\frac{\pi}4\Big )
-
\\
\frac{H_+
(m-1)\sqrt[4]{t^2-(n+N)^2}}{\sqrt{\pi
}\sqrt{(n+N)^2+(m-1)^2(t^2-(n+N)^2)}}\sin\Big(\omega_+ t+\psi_++\frac{\pi}4
\Big).
\label{dotu-pass-0<n<N}
\end{multline}

\begin{remark}	
Two terms in the right-hand side of Eq.~\eqref{dotu-pass-0<n<N}, i.e., the
incident and the scattered wave-fields, are estimated at different moving
points of observation,
e.g., at
\begin{equation}
|n+N|=wt, \quad n>0
\end{equation}
and \eqref{wt_for-n<0}, respectively.  In this sense, formula \eqref{dotu-pass-0<n<N} 
for $n>0$ is not an asymptotically exact one and should be considered as an approximate
solution, whereas the analogous formula 
\eqref{dotu-pass-n<0} 
for $n\leq0$ (written in terms of $w$) has an exact asymptotic
meaning.
\end{remark}

\begin{remark}	
The expressions for the scattered wave-field \eqref{bVNNPass1} and the total
wave-field (Eqs.~\eqref{dotu-pass-n<0} 
or \eqref{dotu-pass-0<n<N})
provide the continuum description of the
corresponding physical processes, i.e., we can assume that $n\in\mathbb R$; see
\cite{Shishkina2023cmat,Gavrilov2022ijhmt}. 
\label{remark-cont-v}
\end{remark}

\begin{remark}	
The first term in the right-hand side of Eq.~\eqref{dotu-pass-0<n<N} (the scattered wave-field) can be considered as the
contribution of the imaginary mirrored source at $n=-N$; see
\cite{Liazhkov2023} where the problem concerning a semi-infinite chain is
considered. {In the
framework of the random (thermal) problem formulated in Sect.~\ref{sec-formulation2},
this mirrored source is not uncorrelated with the real source at $n=N$, as
assumed in \cite{Liazhkov2023}. This leads to the distortion of the analytical solution near
the inhomogeneity, i.e., near the free end of the chain, which is observed,
e.g., in 
Fig.~4 of \cite{Liazhkov2023}.}
\label{remark-L}
\end{remark}

\begin{remark}	
Due to the
coalescence of singular points \cite{Shishkina2023cmat}, asymptotic formulae
\eqref{ins-wave-asymp11},
\eqref{asymp-J_2}
are not valid near the leading wave-fronts: for $w\simeq1$ or, equivalently, for
\begin{equation}
|n-N|\simeq t.
\label{near-leading}
\end{equation}
Analogously,
formula \eqref{asymp-J_2-2} is not valid 
near the leading scattered wave-front 
\begin{equation}
|n+N|\simeq t.
\label{near-ref-leading}
\end{equation}
On the other hand, at the corresponding wave-fronts
Eq.~\eqref{ins-wave-asymp11} is singular,
whereas the transmissive 
\eqref{asymp-J_2-mod}
and the reflected 
\eqref{asymp-J_2-2}
waves are zero.
Therefore, the practical importance of this fact is greater for
Eq.~\eqref{ins-wave-asymp11} describing the leading wave-fronts
\eqref{near-leading} than for 
Eq.~\eqref{asymp-J_2-2}
describing the leading front of the reflected wave \eqref{near-ref-leading}.
The leading front of the transmissive wave described by 
\eqref{asymp-J_2-mod}
coincides with 
the leading left-running wave-front~\eqref{near-leading}.
\label{remark-w1}
\end{remark}
%
%
%
%
%
%
%

\subsection{The contribution from the stop-band}
\label{sect-stop}

Following \cite{Shishkina2023cmat}, we estimate the contribution from the
stop-band at a fixed position $n$.
Due to Eqs.~\eqref{G-def}, 
\eqref{Green-function0-upper},
\eqref{Green-breve},
one has
\begin{equation}
\Isss
=-\frac{\I}{2\pi}\int_{2}^{+\infty}
\frac{\Omega^3(m-1)(-1)^{N+|n|}\EXP{-b(\Omega)(N+|n|)}\EXP{-\I\Omega t}}
{(-\Omega^2+2\EXP{-b(\Omega)}+2)(-m\Omega^2+2\EXP{-b(\Omega)}+2)} \, \d\Omega
.
\label{stop-con}
\end{equation}

The asymptotic order of the contribution from the stop-band depends on the
existence of a mode of oscillation localized near the inclusion at $n=0$.
Such a mode exists if and only if $m<1$ \cite{Montroll1955,Shishkina2023cmat}.
In this case, there
exists a simple root
\begin{equation}
\Omega_0=\frac2{\sqrt{m(2-m)}}
\label{loc-freq-m<1-zero}
\end{equation}
of the second multiplier of the denominator for the
integrand in Eq.~\eqref{stop-con}:
\begin{equation}
-m\Omega_0^2+2\EXP{-b(\Omega_0)}+2=0.
\label{freq-eq-complex-wn-al0}
\end{equation}
The details of how we should understand the
integral in the right-hand side of Eq.~\eqref{stop-con} in this case are
discussed in \cite{Shishkina2023cmat}.
One can get:
\begin{multline}
\Isss
=
-\frac
{(m-1)\Omega_0^3(-1)^{N+|n|}\EXP{-b(\Omega_0)(N+|n|)}
\EXP{-\I \Omega_0 t}}
{-\Omega_0^2+2\EXP{-b(\Omega_0)}+2}
\mathrm{Res}
\Big(
\frac{1}{-m\Omega^2+2\EXP{-b(\Omega)}+2},\, \Omega_0
\Big)
\\+O(t^{-1}).
\end{multline}
Here, the symbol $\Res (f, \Omega_0)$ means the residue of a function $f
(\Omega)$ at a pole $\Omega =\Omega_0$. Calculating the residue yields 
\begin{equation}
\mathrm{Res}
\Big(
\frac{1}{-m\Omega^2+2\EXP{-b(\Omega)}+2},\, \Omega_0
\Big)
=
-\frac{(m-1)\Omega_0}{2(m(m-2)\Omega_0^2+2)}.
\end{equation}
Due to  Eqs.~\eqref{loc-freq-m<1-zero}, \eqref{freq-eq-complex-wn-al0},
one gets:
\begin{equation}
\EXP{-b(\Omega_0)}=\frac{m}{2-m}.
\end{equation}
Taking into account the last expression together with
Eq.~\eqref{loc-freq-m<1-zero}, we derive the following formula:
\begin{equation}
\bVNNStop=
\frac
{2(m-1)(-1)^{N+|n|+1}m^{N+|n|-1}}
{(2-m)^{N+|n|+1}}
\cos(\Omega_0 t)
+O(t^{-1}).
\label{stop-con-c}
\end{equation}

In the case $m>1$, there is no pole in the denominator for the integrand in the
right-hand side of 
Eq.~\eqref{stop-con}, and instead of 
\eqref{stop-con-c},
one gets
\begin{equation}
\bVNNStop=O(t^{-1}).
\end{equation}

Formula 
\eqref{stop-con-c}
describes non-vanishing localized near $n=0$ oscillation.  One can see that
the amplitude of this oscillation exponentially vanishes with $N$. {In this
paper, we assume} that the source at $n=N$ is located far enough from the
inclusion at $n=0$. Thus, we assume, in what follows, that non-vanishing
oscillation $\bVNNStop$ 
has an exponentially small amplitude and can be neglected.

\subsection{Comparison with numerics}
\begin{figure}[htb]  
\centering\includegraphics[width=0.8\textwidth]{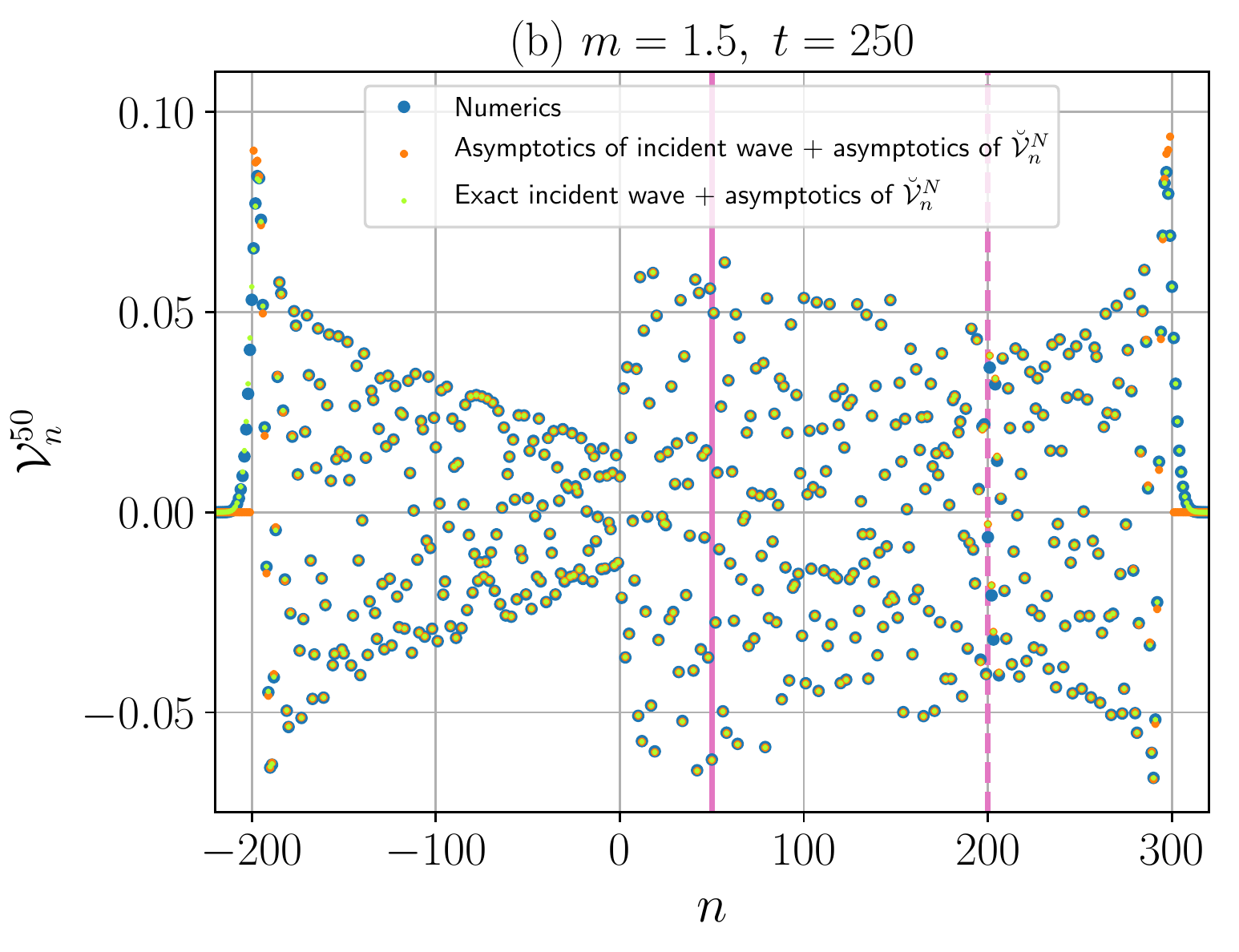}
\centering\includegraphics[width=0.8\textwidth]{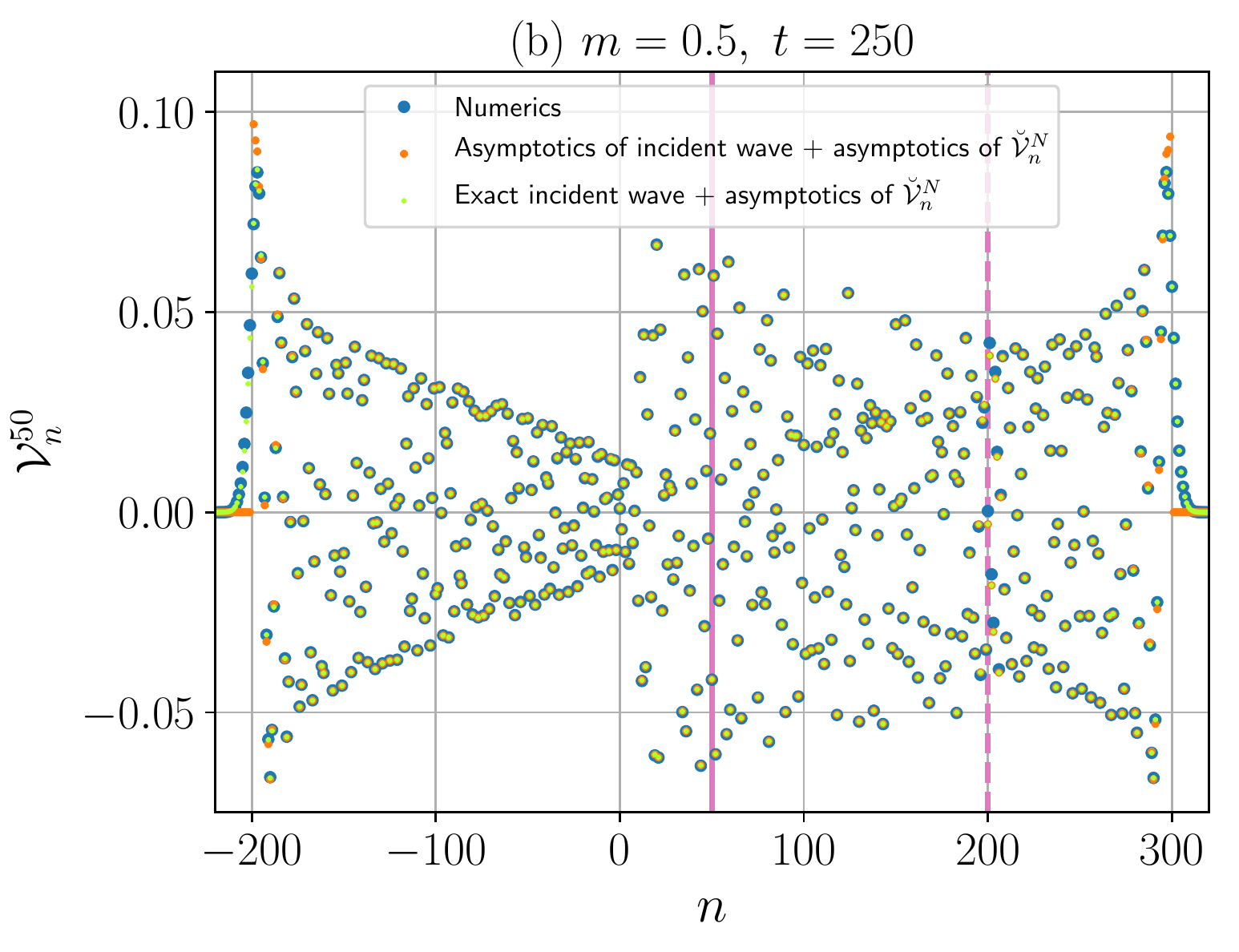}
\caption{The particle velocity $\VF_n^N$ versus the spatial variable $n$. The source position
is indicated
by the vertical magenta solid line. 
The right leading scattered wave-front is indicated
by the vertical magenta dashed line. (a) The case of a heavy defect, (b) the case of a light defect}
\label{V_loc_dif.pdf}
\end{figure}

In Fig.~\ref{V_loc_dif.pdf}, we present 
the spatial plot of the fundamental solution $\VF_n^N(t)\ (n\in \mathbb Z)$
for the particle velocity 
obtained by three different approaches. In the first case, $\VF_n^N(t)$ are found by numerical solution
for system of ordinary differential equations \eqref{chain-eq-basic-al0} 
with periodic boundary conditions \cite{Gavrilov2019cmat}. In the second case, we use 
Eqs.~\eqref{VF-super}, 
\eqref{ins-wave-asymp22} for the incident wave $V_{n-N}$, and 
Eq.~\eqref{bVNNPass1}
($n\leq0$)
or 
Eq.~\eqref{asymp-J_2-2} ($n>0$) 
for the scattered wave $\breve \VF_n^N$.
In the third case, we use the exact expression
for $V_{n-N}$, where $V_n$ is given by Eq.~\eqref{Sro-bessel}
for the incident wave, and the same formulas for the scattered wave as in
the second case. Two sub-plots are presented, which correspond to a heavy
($m>1$) and a light ($m<1$) defect. One can see that all three approaches give
very similar results. The differences can be observed only on the leading incident
wave-fronts, and on the leading scattered wave-front, where the coalescence
of the critical points should be taken into account, see Remark~\ref{remark-w1}.
Furthermore, one can
see that the sub-plots are qualitatively similar, and,
as expected, due to the results of Sect.~\ref{sect-stop} 
for a large enough $N$,
the existence of the localized
mode in the second case (see Sect.~\ref{sect-stop}) does not change the
wave-field.

\afterpage{\clearpage}

\begin{remark}	
Of course, from a pure formal point of view, the localized mode contribution defined by 
Eq.~\eqref{stop-con-c} is the principal part of the solution near the defect.
For very large times, when the propagating component of the wave-field
totally vanishes, the solution near the defect equals the right-hand side of 
Eq.~\eqref{stop-con-c}. However, in this paper, we are mostly interested in
the previous stages of the wave process, when the propagating component still
dominates over the localized one.
\end{remark}

\section{Heat transport: the slow and fast thermal motions}
\label{sect-transport}
In what follows, we consider the {kinetic energy (heat) transport} problem
formulated in Sect.~\ref{sec-formulation2}.
\subsection{$n\leq0$}
Now we calculate the thermal fundamental solution 
\eqref{thermal-fs} taking $\VF_n^N=\VNNPass$ and separate the slow and the
fast
motions following 
\cite{Shishkina2023cmat,Gavrilov2022ijhmt}. 
%
Consider the case $n<0$. Due to Eqs.~\eqref{thermal-fs},
\eqref{dotu-pass-n<0}
\begin{equation}
\TT_n^N
=\frac{2m_nH(1-w)w^2
\cos^2\big(\omega t+\psi+\frac{\pi}4 \big)
}{\pi t\sqrt{1-w^2}(w^2+(m-1)^2(1-w^2))} +o(t^{-1})
=
\Bar{\TT}_n^N+\Hat{\TT}_n^N+o(t^{-1})
,\quad w\not\simeq 1;
\label{kin-pass-n<0}
\end{equation}
where
\begin{multline}
\Bar{\TT}_n^N=
\frac{m_nH(1-w)w^2}{\pi  t \sqrt{1-w^2} \left((m-1)^2 \left(1-w^2\right)+w^2\right)}
\\=
\frac{m_nH_-(n-N)^2}
{
\pi\sqrt{t^2-(n-N)^2}
\big((n-N)^2+(m-1)^2(t^2-(n-N)^2)\big)
}
\label{slow-1}
\end{multline}
is the slow motion, and
\begin{equation}
\Hat{\TT}_n^N=-\Bar{\TT}_n^N\sin2(\omega_- t+\psi_-)
\label{fast-1}
\end{equation}
is the fast motion. 


\subsection{$n>0$}
To calculate the thermal fundamental solution $\TT_n^N$ for $n>0$, we use
approximate solution \eqref{dotu-pass-0<n<N}:
\begin{multline}
\TT_n^N\simeq
\frac{2m_nH_+(m-1)^2\sqrt{t^2-(n+N)^2}\sin^2\big(\omega_+ t+\psi_+ +\frac{\pi}4\big)}
{\pi ((n+N)^2+(m-1)^2(t^2-(n+N)^2))}
+
\frac{2m_nH_-\cos^2\big(\omega_-t+\frac{\pi}4\big )}{\pi\sqrt{t^2-(n-N)^2}}
\\
-
\frac{
4m_nH_+(m-1)\sqrt[4]{t^2-(n+N)^2}
\sin\big(\omega_+t+\psi_+ +\frac{\pi}4\big)\cos\big(\omega_- t+\frac{\pi}4\big )
}
{\pi\sqrt[4]{t^2-(n-N)^2}\sqrt{(n+N)^2+(m-1)^2(t^2-(n+N)^2)}}.
\end{multline}
The last equation can be transformed into the following form:
\begin{equation}
\TT_n^N\simeq
\Bar{\TT}_n^N+\Hat{\TT}_n^N
,
\label{Tn>0}
\end{equation}
where 
\begin{gather}
\Bar{\TT}_n^N=
m_n\Bar T_{n-N}+
\breve{\TT}_n^N
\label{slow-is-sum}
\end{gather}
is the slow motion, 
\begin{gather}
\breve \TT_n^N\=
\frac{H_+m_n(m-1)^2\sqrt{t^2-(n+N)^2}
}
{\pi \big((n+N)^2+(m-1)^2(t^2-(n+N)^2)\big)}
\label{slow-reflected}
\end{gather}
is the scattered component of the slow motion;
\begin{multline}        
\Hat{\TT}_n^N=
m_n\hat T_{n-N}+
\Breve\TT_n^N\sin2(\omega_+ t+\psi_+)
\\
-\frac{2m_nH_+
(m-1)\sqrt[4]{t^2-(n+N)^2}
{
\Big(\cos\big((\omega_++\omega_-)t+\psi_+\big)+\sin\big((\omega_+-\omega_-)t+\psi_+\big)
\Big)
}}
{
\pi\sqrt[4]{t^2-(n-N)^2}\sqrt{(n+N)^2+(m-1)^2(t^2-(n+N)^2)}
}
\label{fast-beat}
\end{multline}
is the fast motion;
$\bar T_{n-N}$ and $\hat T_{n-N}$ are the slow and fast motions in a uniform
chain, defined by Eqs.~\eqref{T-slow-def} and 
\eqref{T-fast-def}, respectively.

\begin{remark}	
The thermal fundamental solution $T_{n-N}$ in a uniform chain, which is defined by 
\eqref{TnN-def},
can be decoupled in the same way 
into the following superposition
\cite{Shishkina2023cmat,Gavrilov2022ijhmt}:
\begin{equation}        
T_{n-N}
=
\Bar T_{n-N}
+
\Hat T_{n-N},
\label{T-pass-def}
\end{equation}
where
\begin{gather}        
\bar T_{n-N}
\=
\frac{H_-}{\pi\sqrt{1-w^2}}=
\frac{H_-}
{\pi\sqrt{t^2-(n-N)^2}
},
\label{T-slow-def}
\\
\Hat T_{n-N}
\=
-
\bar T_{n-N}
\sin 2  \omega_- t 
\label{T-fast-def}
\end{gather}
are the slow and fast motions in the uniform chain, respectively.
\end{remark}
\begin{remark}	
Since $\VF_n^N$ is a continuum quantity (see Remark~\ref{remark-cont-v}), the
slow 
$\Bar{\TT}_n^N$
and the fast 
$\Hat{\TT}_n^N$
motions can be naturally continualized in the range $|n|\geq1$, where
$m_n=1$, see Remark~12 in \cite{Shishkina2023cmat}. 
\label{remark-cont}
\end{remark}

Thus, the slow motion is the thermal fundamental solution averaged over the
fast phases, namely, over 
\begin{equation}
\varphi_0\=2(\omega_- t+\psi_-)
\end{equation}
for $n\leq0$;
\begin{equation}
\varphi_1\=2(\omega_+t+\psi_+),\quad 
\varphi_2\=2\omega_-t,\quad
\varphi_3\=(\omega_++\omega_-)t+\psi_+,\quad
\varphi_4\=(\omega_+-\omega_-)t+\psi_+ 
\label{varphi14}
\end{equation}
for $n>0$. 
\begin{figure}[htb]  
\centering\includegraphics[width=0.65\textwidth]{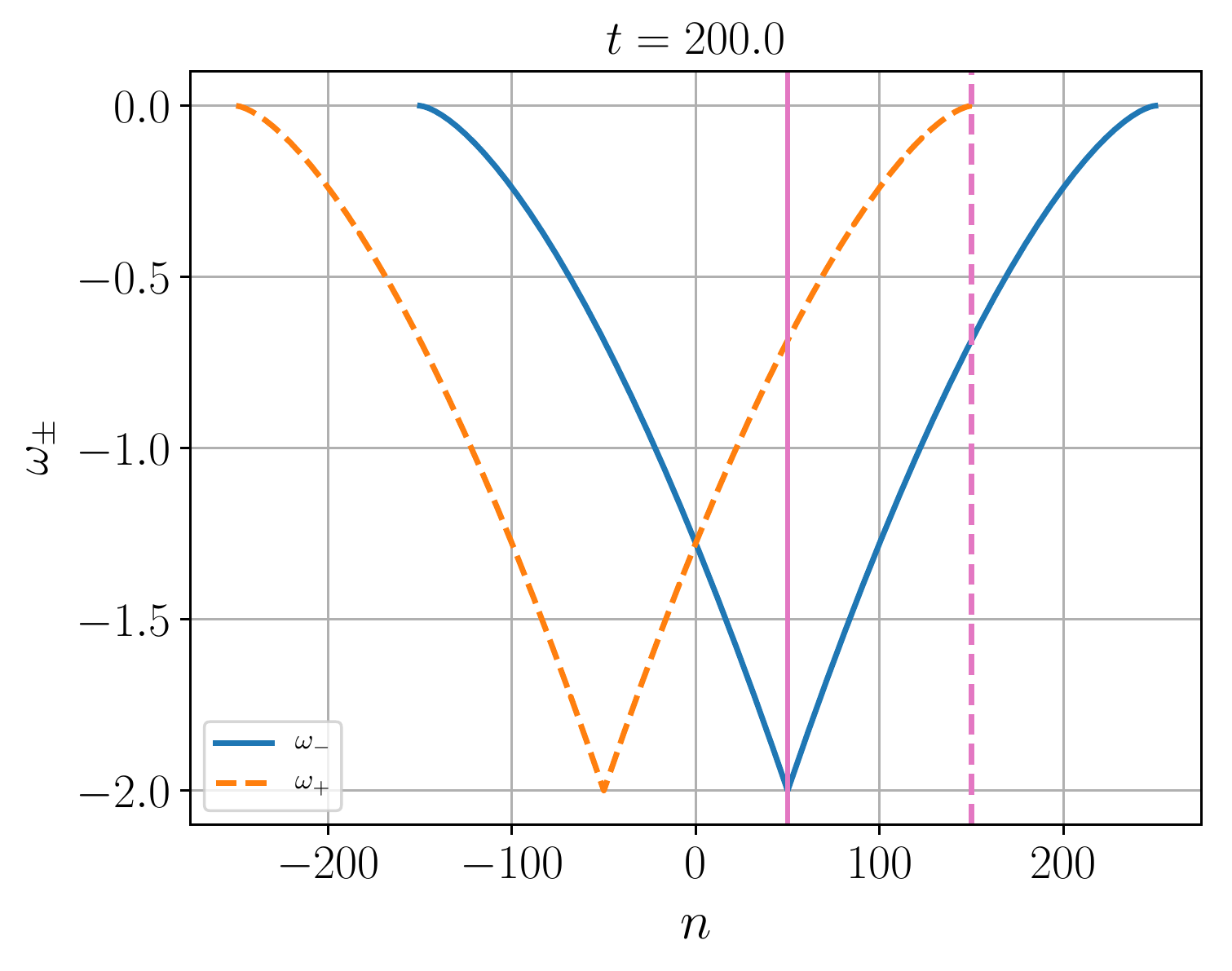}
\caption{Frequencies $\omega_-$ and $\omega_+$ versus $n$.
The source position
is indicated
by the vertical magenta solid line.
The right leading reflected wave front is indicated
by the vertical magenta dashed line}
\label{omegas.pdf}
\end{figure}
\begin{remark}	
In Fig.~\ref{omegas.pdf} one can see
the plots of frequencies $\omega_-$ and $\omega_+$ versus $n$. 
One can observe that at the leading wave-fronts
\eqref{near-leading}
one of the fast phases $\varphi_0$ or $\varphi_2$ transforms to a slow
quantity (for $n<0$
and $n>0$, respectively). In principle, this breaks the slow-and-fast
decoupling, but 
it is not a big problem for us since asymptotics 
\eqref{ins-wave-asymp11} is in any case wrong at these wave-fronts (see
Remark~\ref{remark-w1}). 
For $n>0$, at the leading reflected wave-front
\eqref{near-ref-leading},
the phase $\varphi_1$ transforms to a slow quantity. This also
corresponds to a domain where approximation 
\eqref{asymp-J_2-2}
is not valid (see Remark~\ref{remark-w1}).  Additionally, the phase 
$\varphi_4$ transforms to a slow quantity for
$n\to+0,\ n\in\mathbb R$. This breaks at $n\to+0$ the slow-and-fast decoupling for 
$\TT_n^N$, which is proportional to $(\VF_n^N)^2$:
the fast motions caused by the incident and reflected waves originate a slow
motion. The last observation will be illustrated in what follows; see
Sect.~\ref{sec-compare}.
\label{remark-beat}
\end{remark}

\subsection{Comparison with numerics}
\label{sec-compare}

\begin{figure}[htb]  
\centering\includegraphics[width=0.8\textwidth]{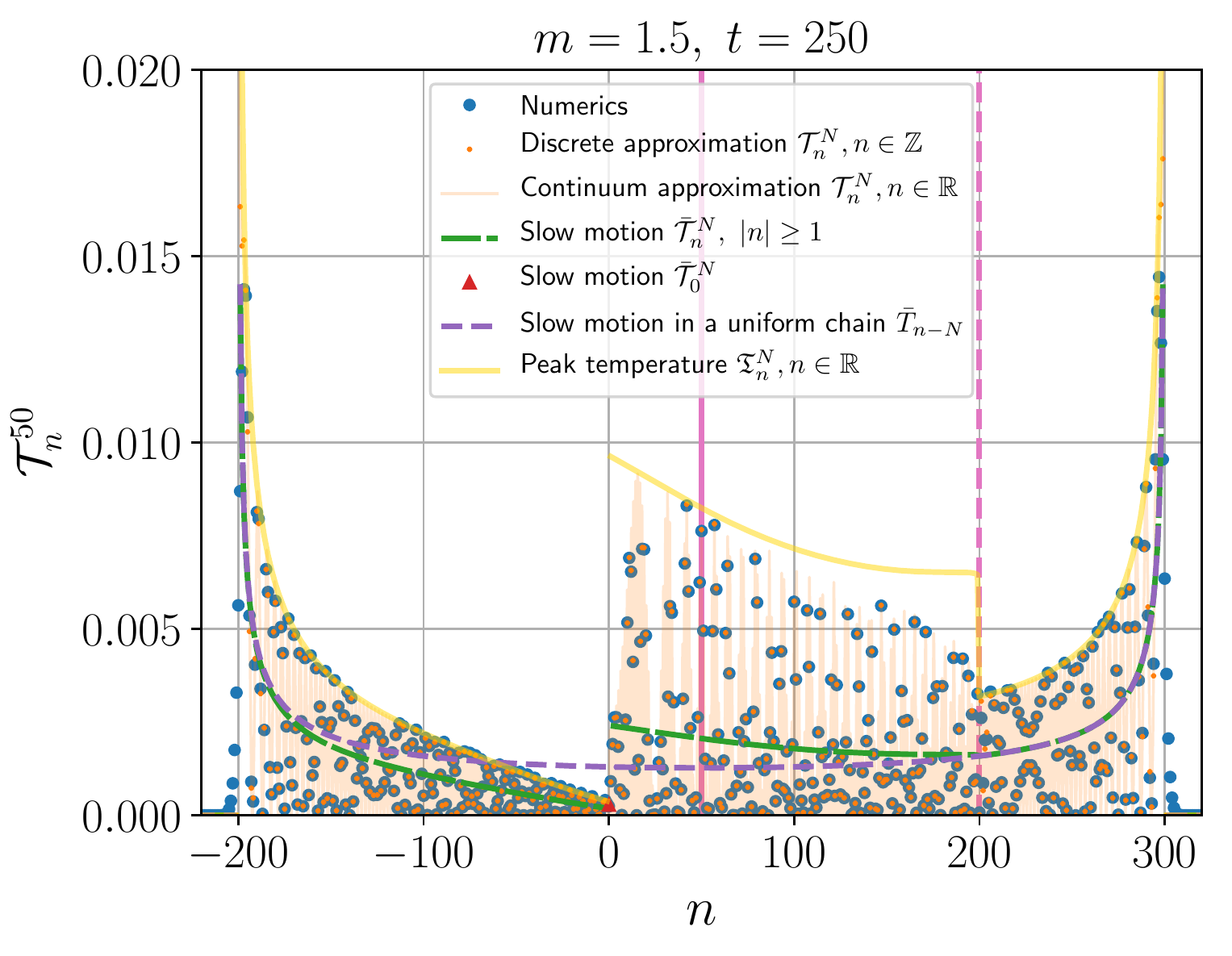}
\caption{Kinetic temperature $\TT_n^N$ versus spatial variable $n$. The source position
is indicated
by the vertical magenta solid line.
The right leading reflected wave-front is indicated
by the vertical magenta dashed line}
\label{Ts_dif.pdf}
\end{figure}

In Fig.~\ref{Ts_dif.pdf}, we present 
a spatial plot of the fundamental solution for the kinetic temperature $\TT_n^N$ expressed by 
Eq.~\eqref{thermal-fs}
and obtained by two approaches. In the framework of the first approach,
particle velocities $\VF_n^N(t)$ 
in Eq.~\eqref{thermal-fs}
are found numerically solving
system of ordinary differential equations \eqref{chain-eq-basic-al0} 
with periodic boundary conditions \cite{Gavrilov2019cmat}.
In the second case, we use the obtained above
analytic approximations for the fundamental solution 
$\VF_n^N(t)$, which lead to
Eqs.~\eqref{kin-pass-n<0}--\eqref{fast-1} (for $n\leq0$) or 
Eqs.~\eqref{Tn>0}--\eqref{T-fast-def} (for $n>0$). Here, $n\in \mathbb Z$ or 
$n\in \mathbb R$
for the discrete approximation or
for the continuum one, respectively. One can see that discrete approximation is in 
excellent agreement with numerics everywhere except near 
leading wave-fronts \eqref{near-leading}, where asymptotics 
\eqref{ins-wave-asymp11} is not valid (see Remark~\ref{remark-w1}). The
agreement is a bit worse near the reflected wave leading front 
\eqref{near-ref-leading} (see Remark~\ref{remark-w1} again).

The slow motion $\bar \TT_n^N$, which is also plotted in 
Fig.~\ref{Ts_dif.pdf}, 
has essentially different limiting values
at the boundaries for domains of continuity $n=\pm1$, $n\in\mathbb R$. 
It is minimal (and close to zero) for small
non-positive~$n$, where a thermal shadow behind the defect is forming. For
small positive values of $n$, the slow motion is essentially larger.
Thus, at the
defect, one can observe a pronounced jump discontinuity of the slow kinetic temperature.

One can see that the slow motion looks like
a spatial average for $\TT_n^N$. 
In the ranges 
where the fast motion contains a unique harmonic, we can introduce ``a peak
temperature''
\begin{equation}
\mathfrak T_n^N\=2\bar \TT_n^N,\quad n\in\mathbb R;
\quad n\leq0\quad \text{or}\quad n>t-N.
\end{equation} 
One can see that the plot
for $\mathfrak T_n^N$ looks like the envelope for $\bar \TT_n^N,\ n\in \mathbb
R$ in these intervals.
In
the range $0<n<t-N$, where the fast motion consists of multiple harmonics, see
Eq.~\eqref{fast-beat}, {a beat} is observed since one of a number of the
fast phases  
becomes a slow quantity for
$n\to+0,\ n\in\mathbb R$;
see Remark~\ref{remark-beat}. 
Introducing the peak
temperature as 
\begin{equation}
\mathfrak T_n^N\=4\bar \TT_n^N,\quad n\in\mathbb R;
\quad 0<n<t-N,
\end{equation}
one can see that for small enough positive $n$, the plot for  $\mathfrak T_n^N$ 
again looks like the envelope for $\bar \TT_n^N$, $n\in \mathbb R$. Thus,
{the value of the 
slow temperature for small $n>0$ is close to 
a quarter of the ``nearest global maximum value\footnote{In a certain
large enough neighbourhood of zero.}'' for $\TT_n^N$,
$n\in\mathbb R$, which is observed in Fig.~\ref{Ts_dif.pdf} at
$n\approx15.6$.}

\begin{figure}[htb]  
\centering\includegraphics[width=0.65\textwidth]{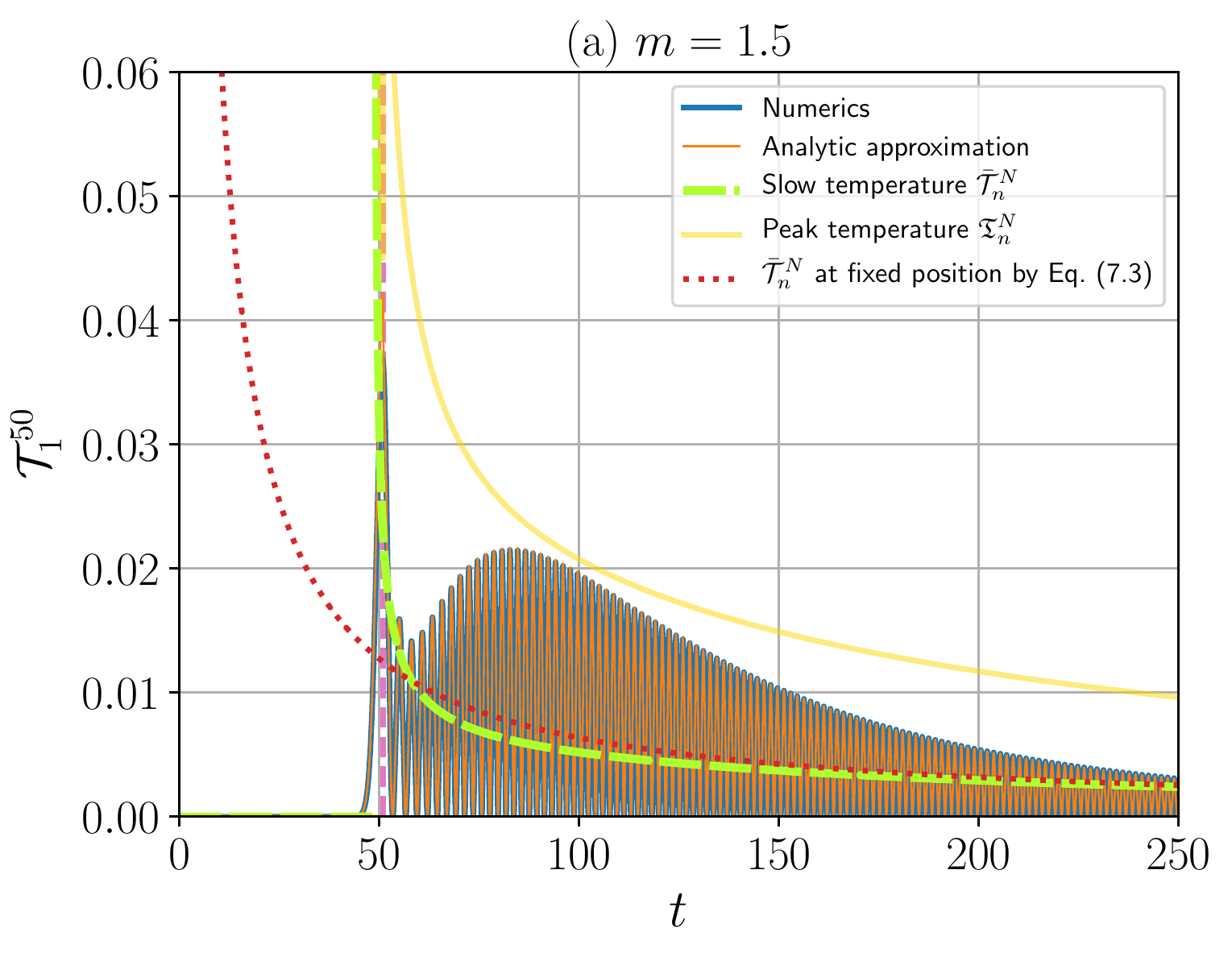}
\centering\includegraphics[width=0.65\textwidth]{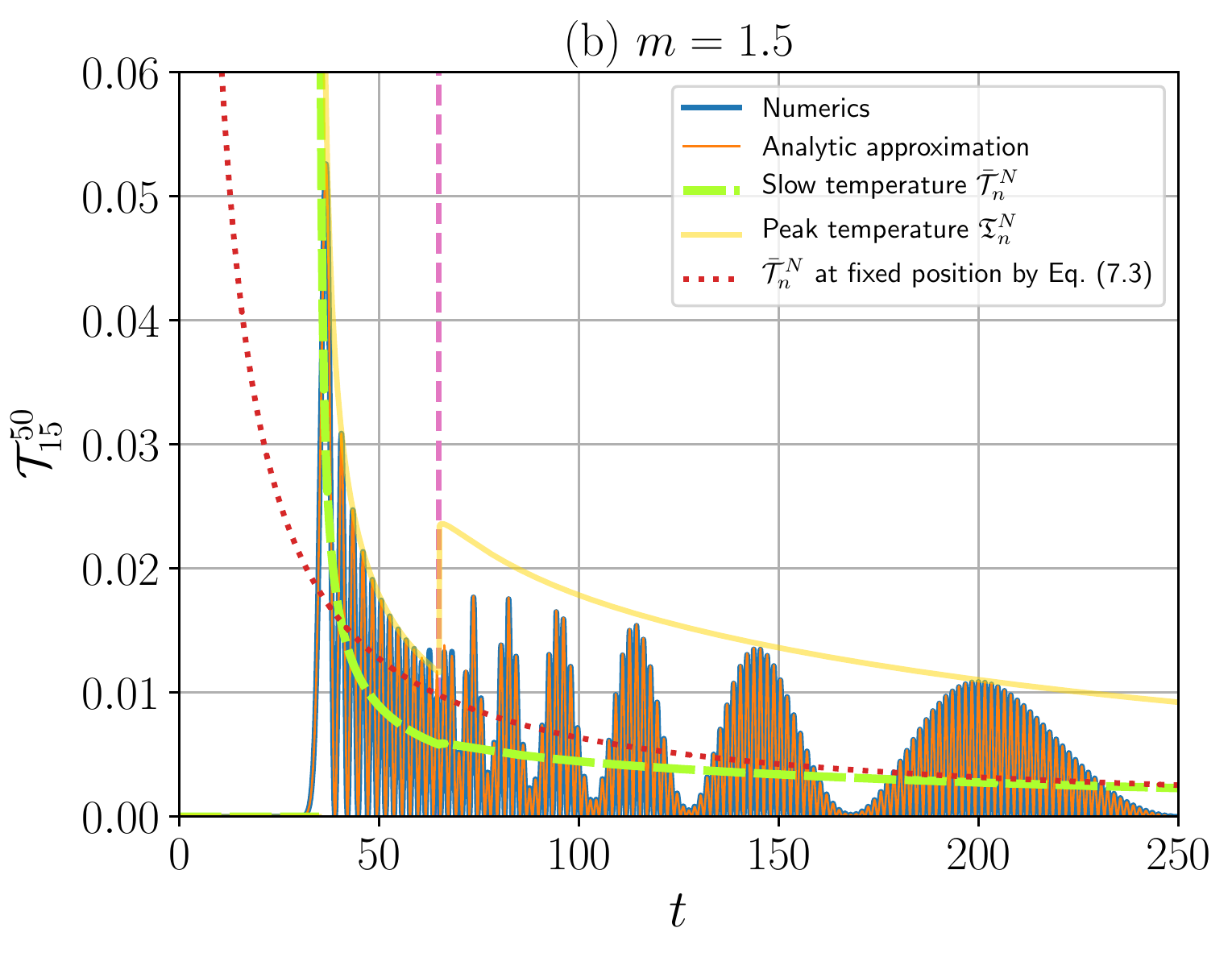}
\centering\includegraphics[width=0.65\textwidth]{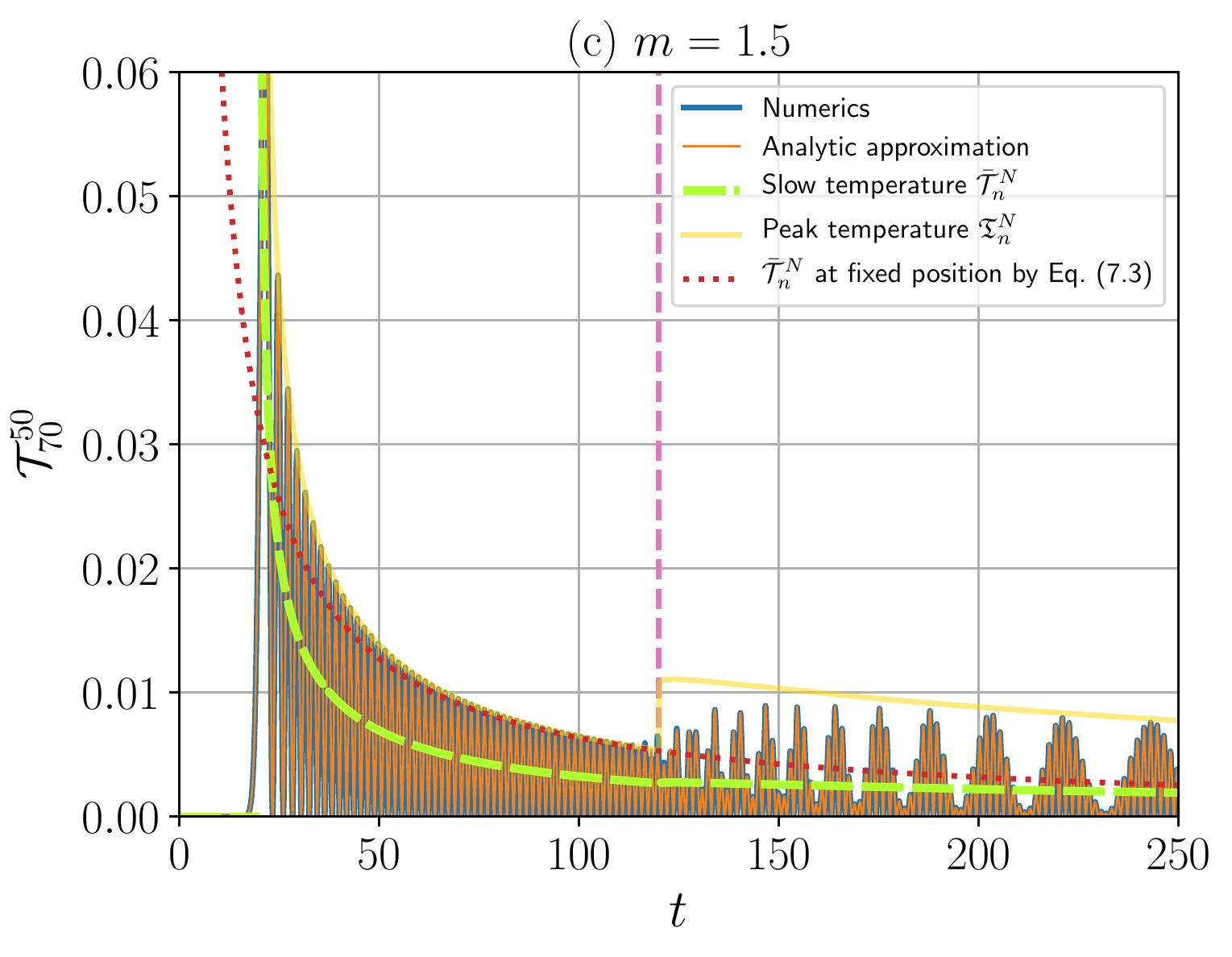}
\caption{Kinetic temperature $\TT_n^N$ versus time $t$ for various positive $n$.
The instant of coming of the leading reflected wave-front is indicated
by the vertical magenta dashed line}
\label{Tt_dif.pdf}
\end{figure}

In Fig.~\ref{Tt_dif.pdf}, we present 
temporal plots
of the fundamental solution for the kinetic temperature $\TT_n^N$ 
obtained for several values of $n>0$ 
by the same two approaches (numeric and
analytic ones) as we used for Fig.~\ref{Ts_dif.pdf}.
Again, in the range 
{$N-n<t<N+n$}, where
the fast motion contains a unique harmonic,
the plot
for $\mathfrak T_n^N$ is the envelope for $\TT_n^N$. 
For larger times $t>N+n$ the peak temperature 
$\mathfrak T_n^N$ again becomes the envelope\footnote{Or, better to say, the envelope for the envelope.}
for $\TT_n^N$ after some
transient process. The beat period becomes infinity for $n\to+0$; therefore, the
approximate description of the energy transport by the slow motion becomes a bit doubtful
for small positive $n$ (see Fig.~\ref{Tt_dif.pdf}(a)). 
For larger $n$, this quantity 
again acquires a clear physical meaning (see Fig.~\ref{Tt_dif.pdf}(b)--(c)),
{though the beat period grows and approaches infinity as $t\to\infty$}.

\begin{figure}[htb]  
\centering\includegraphics[width=0.65\textwidth]{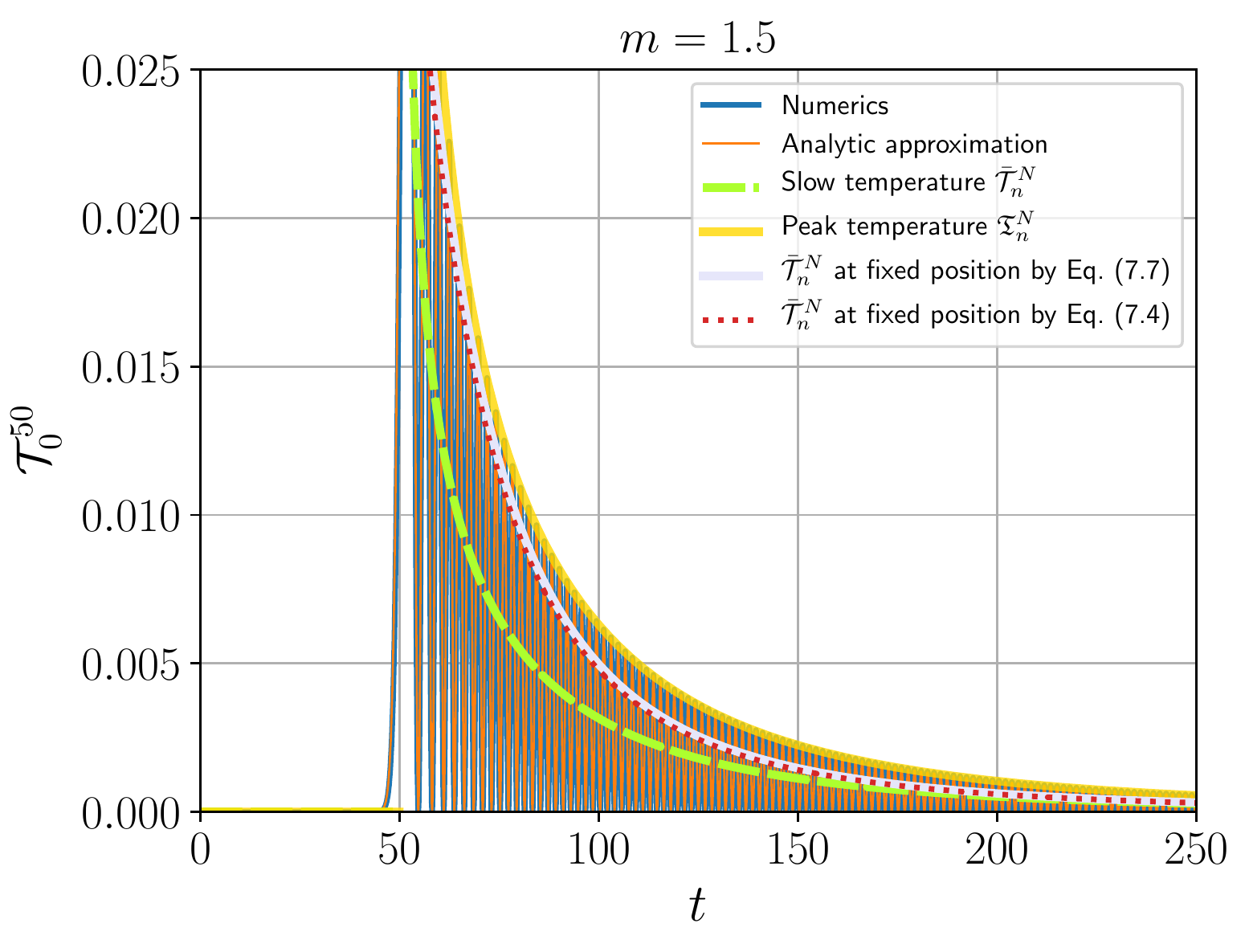}
\caption{Kinetic temperature $\TT_n^N$ versus time $t$ ($n=0$)}
\label{Tt0_dif.pdf}
\end{figure}

In Fig.~\ref{Tt0_dif.pdf}, we present 
the temporal plot 
of the fundamental solution $\TT_n^N$
obtained for $n=0$. 
One can see that the fast motion
contains a unique harmonic, and 
the plot
for $\mathfrak T_n^N$ is the envelope for $\TT_n^N$
as expected. The plots for various $n<0$ have the same qualitative character
as for $n=0$,
and, thus, we do not demonstrate more ones.

\afterpage{\clearpage}

\section{Anti-localization for $n\leq0$: a thermal shadow behind the defect,
the Kapitza thermal resistance}

\label{sec-anti}

In Sect.~\ref{sec-compare}, we have shown numerically that the slow
temperature $\bar\TT_n^N$ quickly changes near the defect. Thus, we observe
the non-stationary analogue of the effect characterized by the Kapitza thermal resistance \cite{Kapitza1941}. Note
that in \cite{Paul2020,Gendelman2021} a similar model is used
to describe the Kapitza thermal resistance in the case where two stationary
thermal sources are located at the ends of a finite chain with an isotopic
defect. 

Let us formally calculate for a fixed $n>0$ and $n\leq0$ the large time asymptotics of the components
$m_n\bar T_{n-N}$ and $\breve\TT^N_n$ in the right-hand side of 
Eq.~\eqref{slow-is-sum}
of the slow temperature $\bar\TT^N_n$.

For $n>0,\ n\in\mathbb Z$ (or $n\geq1,\ n\in\mathbb R$ if we discuss
continuum quantities),
according to 
\eqref{TnN-def},
\eqref{slow-reflected},
{we have: $m_n=1$,} 
\begin{gather}
m_n\bar {T}_{n-N}=\frac{1}{\pi  t}
+O(t^{-3}),
\label{incident-fixed-as}
\\
\breve {\TT}_n^N=\frac{1}{\pi  t}
+O(t^{-3}),
\end{gather}
and, thus, 
\begin{gather}
\Bar{\TT}_n^N=\frac{2}{\pi  t}
+O(t^{-3}).
\label{sum-fixed-as}
\end{gather}
Formulae~\eqref{incident-fixed-as}--\eqref{sum-fixed-as}
are clearly derived in a not entirely correct way, i.e., {by calculating 
the asymptotics at a fixed position of the solution obtained at a moving point of
observation.} They demonstrate that reflection from the defect results in the doubling
of the slow temperature $\bar\TT_n^N$ for all $n>0$. In Fig.~\ref{Tt_dif.pdf}, one can see
that numerical calculations confirm such a conclusion for very large times:
{the red dotted curves asymptotically approach the corresponding green dashed ones} in all plots.
According to Eqs.~\eqref{incident-fixed-as}--\eqref{sum-fixed-as}, the slow
temperature at this final stage does not depend on $n$ for $n>0$.

For $n\leq0$, according to Eq.~\eqref{slow-1}, one obtains:
\begin{equation}
\Bar{\TT}_n^N=\frac{m_n(n-N)^2}{\pi  (m-1)^2 t^3}
+O(t^{-4}).
\label{T-moving-fixed-pos}
\end{equation}
\begin{remark}  
\label{remark-incorrect}
The result for $n\leq0$,  generally speaking,  is obtained incorrectly since we have
dropped the terms 
of order $o(t^{-1/2})$ when
calculating  
the asymptotics 
\eqref{dotu-pass-n<0}
of the particle velocity $\VNNPass$
at a moving point of observation. These terms correspond
to the terms of order $o(t^{-1})$ in expansions for $\TT_n^N$, whereas the
principal term in 
\eqref{T-moving-fixed-pos}
is of order $t^{-3}$.
However, one
can see in Fig.~\ref{Tt0_dif.pdf} that formula \eqref{T-moving-fixed-pos}
describes the final stage of
evolution for $\TT_n^N$ quite well. To verify Eq.~\eqref{T-moving-fixed-pos}, 
in Sect.~\ref{sect-asy-fixed}, we construct
the large time asymptotics for the particle velocity
$
\VF_n^N
$
and the slow temperature 
$\Bar{\TT}_n^N$
at a fixed position $n\leq0$.
\end{remark}
%
%
%


\subsection{Asymptotics for the particle velocity and the slow temperature at
a fixed position $n\leq0$}
\label{sect-asy-fixed}

To calculate the fixed position 
asymptotics, we deal with integral representation~\eqref{V-1whole}
and apply the method of stationary phase. The only critical point is the cut-off
frequency; thus, we need to calculate the corresponding contribution for integrals
over the pass-band and the stop-band. Take $n\leq0$.
We base on the expansions for the amplitude
\begin{equation}
 C_n^N\=\Omega \mathcal G_n^N
\label{C-def}
\end{equation}
in the pass-band and the stop-band,
see Eqs.~\eqref{Cpass-expansion},
\eqref{Cstop-expansion} which are obtained in
Appendices~\ref{app-pass} and \ref{app-stop}, respectively.
Now, the total contribution from the cut-off frequency 
can be
calculated in the same way as in Sect.~7.3 of \cite{Shishkina2023cmat}, 
where the corresponding expansions of the amplitude are given by
Eqs.~(7.13) and (7.49), respectively. The first (zero order) terms in
the right-hand sides of Eqs.~\eqref{Cpass-expansion}, \eqref{Cstop-expansion} 
are equal to each other, and
the corresponding total contribution from the pass-band and the stop-band is
zero. The second (square root) terms 
equal the corresponding terms in the right-hand side of Eqs.~(7.13),
(7.49) in \cite{Shishkina2023cmat} with accuracy to the multiplier 
$(-1)^{N-n}  \big(1+2 (N-n)(m-1)\big)$.
Thus, one gets
(see Eqs.~(7.54) and (10.6) in \cite{Shishkina2023cmat}, respectively):
\begin{gather}
\VF_n^N=
\frac{(-1)^{N-n} \big (1+2 (\k)(m-1)\big)\sin\big(2t-\frac{\pi}4\big)}{2\sqrt\pi(m-1)^2t^{3/2}}
+o(t^{-3/2}),
\label{V-fixed-pos}
\\
\bar{\mathcal T}_n^N=\frac{m_n\big(1+2 (\k)(m-1)\big)^2}{4\pi(m-1)^4t^3}+o(t^{-3}).
\label{T-fixed-pos}
\end{gather}
Provided that 
\begin{equation}
2(N-n)|m-1|\gg1,
\end{equation}
i.e., number $N$ is large enough\footnote{This is assumed in our paper, see 
Sect.~\ref{sect-stop}.},  
the right-hand sides of Eqs.~\eqref{T-moving-fixed-pos} and 
\eqref{T-fixed-pos} almost coincide (see Fig.~\ref{Tt0_dif.pdf}). Thus,
Eq.~\eqref{T-moving-fixed-pos} really provides ``an almost correct result''.



\subsection{The non-stationary transmission function for the kinetic temperature}
\begin{figure}[htb]  
\centering\includegraphics[width=0.65\textwidth]{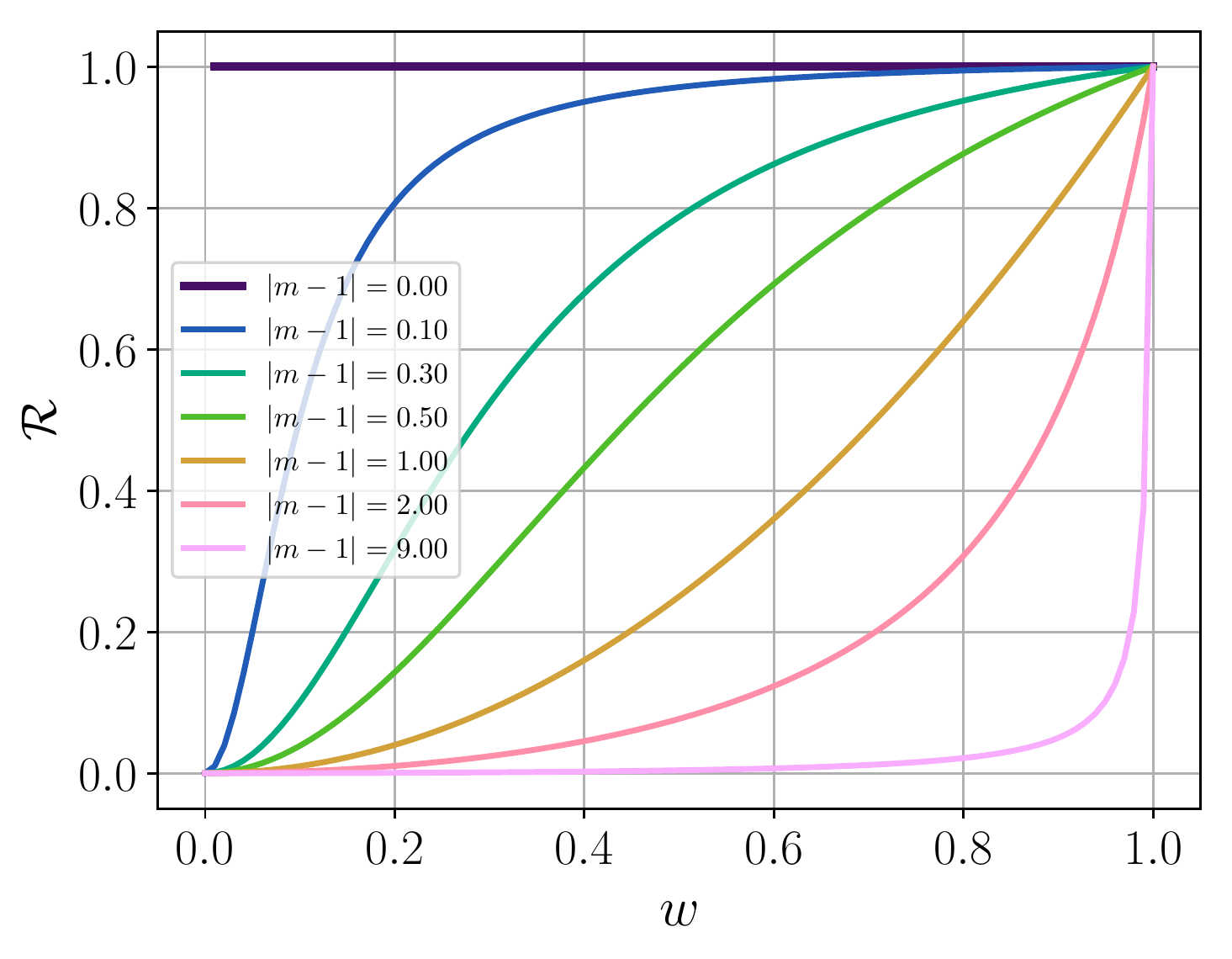}
\caption{The non-stationary transmission function for the kinetic temperature
$\mathscr R$ versus $w$}
\label{ratio-R.pdf}
\end{figure}

The obtained Eqs.~\eqref{V-fixed-pos}, \eqref{T-fixed-pos}
show that for all $n\leq0$, we observe the phenomenon of
anti-localization of non-stationary quasi-waves. This is the zeroing of the non-localized propagating component of
the wave-field in a neighbourhood of an inclusion 
\cite{Shishkina2023cmat,Shishkina2023jsv}. This neighbourhood
expands with time, capturing more and more material points of the chain.
However, this process never leads to a total blocking \cite{Glushkov2006,Glushkov2006a} 
of energy propagation toward
$n\to-\infty$ but only to some distortion of the incident wave-field. Indeed, in the interval
$n\leq0$,
consider the ratio 
\begin{equation}
\mathscr R(w)=
\frac
{\Bar{\TT}_{n}^N}{\bar T_{n-N}}=\frac{w^2}{(m-1)^2 \left(1-w^2\right)+w^2}
\label{R-measure}
\end{equation}
as the function of the speed $w$ defined by 
Eq.~\eqref{wt_for-n<0},
which we call the non-stationary transmission
function for the kinetic temperature.
Here 
${\Bar{\TT}_{n}^N}$ and $\bar T_{n-N}$ are calculated by 
Eqs.~\eqref{slow-1} and \eqref{T-slow-def}, respectively.
The plot of ratio $\mathcal R$ versus $w$ is presented
in Fig.~\ref{ratio-R.pdf}.
One can see that the greater the quantity $|m-1|$,
the wider the anti-localization zone (i.e., the thermal shadow) behind the defect, 
and more wave energy concentrates closer to the leading wave-front $w=1$.
For $m\neq1$, any material particle of the chain with $n\leq0$ 
after some time from the instant of passing the leading wave-front moving at $w=1$ enters
the thermal shadow, where the kinetic temperature
decreases according to Eq.~\eqref{T-fixed-pos} as $O(t^{-3})$.

\begin{remark}  
The cold point at the defect (see Fig.~8(a) in \cite{Gendelman2021}), apparently, is the only
anti-localization point observed in the system with two thermal sources at both sides
of an isotopic defect.

\end{remark}

\begin{remark}  
\label{remark-anti}
The anti-localization of non-stationary waves was introduced previously in the
framework of the problems \cite{Gavrilov2023DD,Shishkina2023cmat,Shishkina2023jsv}, where a loading is applied to a defect. In this
paper, we show for the first time how the anti-localization influences a wave
process when a loading and a defect are located at different
positions.
\end{remark}
%

\section{Conclusion}

In the paper, we have considered two problems. The first one is the
deterministic problem concerning elastic wave scattering on an isotopic defect
in a one-dimensional harmonic crystal (a linear chain).  The chain is
subjected to a unit impulse point loading applied to a particle far enough from
the defect. In the framework of the second problem, we assume that applied
point impulse excitation has random amplitude. This allows one to model the
heat transport in the chain and across the defect as the transport of the
mathematical expectation for the kinetic energy and to use the conception of
the kinetic temperature. 


The most important result related to the first (scattering) problem is 
Eqs.~\eqref{bVNNPass1}, 
\eqref{asymp-J_2-2}, which provide for $t\to\infty$ a continuum approximation
for the fundamental solution
describing the scattered wave-field in the time domain. This approximation is obtained
as a large time asymptotics at
a moving point of observation, and it is in excellent agreement with
corresponding numerical calculations (see Fig.~\ref{V_loc_dif.pdf}).
We have shown that in the case of loading applied far enough from the defect,
which is assumed in the paper, the scattered wave-fields caused by heavy
($m>1$) and light ($m<1$) defects are qualitatively similar. The localized
non-vanishing in time oscillation, which exists in the system in the case $m<1$
\cite{Montroll1955,Shishkina2023cmat}, has exponentially vanishing with 
the
distance (between the defect and the point of the loading)
amplitude;
see Eq.~\eqref{stop-con-c}. Thus, the localized oscillation can be neglected.
This is an essential difference from the case when the
loading is applied at the defect considered in our previous paper
\cite{Shishkina2023cmat}.
In the latter case, the wave-field caused by a light defect 
contains a pronounced non-vanishing localized component. 

It is interesting that the choice of the moving point of observation, which we use
to get the corresponding asymptotics is generally ambiguous. In previous
studies \cite{Shishkina2023cmat,Shishkina2023jsv,Gavrilov2023DD},
we considered the case when the loading was applied to the
defect. We followed the natural choice, which was to consider the fronts 
emerging
at the instance of the impulse loading and then moving
at a constant speed away from the loaded particle, i.e., from the defect. Now, we
consider the problem, where the loading and the defect are in the different
positions, and it is not clear what is ``a natural choice''. We have tried
to use various choices (see Appendix~\ref{app-C}) and obtained different formally
correct asymptotics. We have shown that the only one among the proposed
asymptotics is applicable as an approximate solution, which characterizes the wave-field in terms of the particle number $n$ co-ordinate and time $t$.

The most important result related to the second (heat transport) problem is 
Eqs.~\eqref{slow-1}, \eqref{slow-reflected}, which provide for $t\to\infty$ 
the expressions for the slow (in time) component of the kinetic temperature in the
chain, i.e., the slow motion. To intuitively understand what the slow
motion is, the reader can look through
Figs.~\ref{Ts_dif.pdf}--\ref{Tt0_dif.pdf}.
In discrete harmonic systems,
where a stochastic loading is distributed in space
\cite{krivtsov2015heat,krivtsov-da70,Sokolov2021,Kuzkin2017fast,Kuzkin2019} or in time
\cite{Gavrilov2019cmat,Gavrilov2022cmat,Gavrilov2020cmat}, according to
numerical calculations,
the fast component vanishes. Thus, we expect that to describe the heat
transport for such a loading, 
it is enough to calculate a convolution of the distributed loading with  
the fundamental solution for the slow motion without any spatial
or temporal averaging.
However, in the range $n>0$, especially for small~$n$,
such a description is perhaps too oversimplified; see 
Remark~\ref{remark-beat} and Sect.~\ref{sec-compare}.

The obtained solution allows us to show
that there is a thermal shadow behind the defect: the order of
vanishing for the slow temperature is larger for the particles behind the
defect than for the particles between the loading and the defect 
(see Eqs.~\eqref{sum-fixed-as},
\eqref{T-moving-fixed-pos} and Fig.~\ref{Ts_dif.pdf}). 
From a pure mathematical point of view, such a
result cannot be obtained correctly based on the asymptotic solution at a moving
point of observation
(see Remark~\ref{remark-incorrect}).
Therefore, we have verified and confirmed it by considering asymptotics at a
fixed position (see Sect.~\ref{sect-asy-fixed}). The asymptotics at a
fixed position describes only the last stage of the evolution of corresponding
quantities, i.e., the particle velocity or the kinetic temperature, whereas
the asymptotics at a moving point of observation,
suggested in Sect.~\ref{sect-n<0}, describes 
earlier stages (see Figs.~\ref{Tt_dif.pdf}, \ref{Tt0_dif.pdf}).
Due to the presence of the shadow, the continuum slow kinetic temperature 
has a jump discontinuity at the defect. Thus, the system under
consideration can be a simple model for the non-stationary phenomenon
analogous to one characterized by the Kapitza thermal resistance \cite{Gendelman2021,Kapitza1941,Lumpkin1978,Paul2020}.

The presence of the thermal shadow is related to a non-stationary wave phenomenon,
which we call the anti-localization of non-stationary waves.
This is the zeroing of the non-localized propagating component of
the wave-field in a neighbourhood of an inclusion 
\cite{Shishkina2023cmat,Shishkina2023jsv}. 
In previous studies 
\cite{Shishkina2023cmat,Shishkina2023jsv},
where the anti-localization was introduced, the case when the loading and the
defect were at the same position was considered
(see Remark~\ref{remark-anti}). In the latter case, the zeroing is
observed in a two-sided neighbourhood of the defect. In this
paper, the anti-localization is observed only for the defect and particles behind
the defect with respect to the point of loading ($n\leq0$).
This one-sided neighbourhood
expands with time, capturing more and more material points of the chain.
However, this process never leads to a total blocking  
of energy propagation toward
$n\to-\infty$ (as it is observed,
e.g., in \cite{Glushkov2006,Glushkov2006a}), but only to a distortion of the
incident wave-field due to the defect. We propose 
as a measure for this distortion the non-stationary transmission function for
the kinetic temperature, which can be calculated by 
Eq.~\eqref{R-measure}. One can see 
in Fig.~\ref{ratio-R.pdf}
that the greater the quantity
$|m-1|$ (the absolute value of the difference between the defect mass
and the mass of the regular particle), the more distortion of the wave-field is  
observed at a far zone behind the defect.

We suppose that, without essential modifications, the methodology proposed in
this paper can be applied to other similar problems concerning more
complicated one-dimensional harmonic crystals and other types of defects
(or interfaces).

\section*{Acknowledgements}
The authors are grateful to A.P.~Kiselev, A.M.~Kriv\-tsov, V.A.~Kuzkin,
S.D.~Liazhkov, Yu.A.~Mochalova
for useful and stimulating discussions.

\section*{Funding}
This work is supported by the Russian Science Foundation (project 22-11-00338).

\appendix

\section{Asymptotics at alternative moving points of observation}
\label{app-C}
The choice of the moving observation fronts is generally ambiguous. One can try to
construct asymptotics on moving observation points different from 
Eq.~\eqref{wt_for-n<0}.
Consider the case $n\leq0$ (the scattered wave for $n>0$ as 
in Sect~\ref{difn>0}
can be
calculated using the evenness of $\bVNNPass$ {with respect to $n$}). Let us estimate the integral $\Idva$ on a moving point of
observation different from the one given by Eq.~\eqref{wt_for-n<0}.  We can
try to use the following moving point of observation instead of Eq.~\eqref{wt_for-n<0}:
\begin{equation}
|n|=wt.
\label{as1-n}
\end{equation}
The integral $\Idva$  
\eqref{I-pass-left-pre}
can be represented in the form of
Eq.~\eqref{I-pass-left}, where
\begin{equation}
A^{\mathrm{pass}}(\Omega)=\frac{(m-1)\Omega \EXP{\I N \arccos \frac{2-\Omega^2}2} }{-(4-\Omega^2)+\I(m-1)\Omega\sqrt{4-\Omega^2}},
\end{equation}
$\phi(\Omega)$ is given by Eq.~\eqref{phase-phi(Omega)}.
Applying the procedure of the method of stationary phase in the same way
as it is done in Sect.~\ref{sect-n<0}, instead of Eq.~\eqref{asymp-J_2} one 
gets:
%
\begin{multline}
(\VFF_n^N)^{\mathrm{pass}}=
-\frac{H(1-w)(m-1)\sqrt[4]{1-w^2}}{\sqrt{\pi t}\big(w^2+(m-1)^2(1-w^2)\big)}
\Big(
\Big((m-1)\sqrt{1-w^2}\cos \FFF+w\sin \FFF\Big)
\cos\Big(\omega t+\frac{\pi}4\Big )
\\
+
\Big(w\cos \FFF -(m-1)\sqrt{1-w^2}\sin \FFF \Big)
\sin\Big (\omega t+\frac{\pi}4 \Big ) 
\Big)
+O(t^{-1}).
\label{asymp-J_2-1}
\end{multline}
Here
\begin{gather}
F=F_0\arccos (2w^2-1),
\label{F=F0}
\\
F_0=N.
\label{F=N}
\end{gather}
Formula 
\eqref{asymp-J_2-1} can be transformed into the form of 
Eq.~\eqref{asymp-J_2-mod}, wherein we should substitute 
\begin{equation}
\psi_\ast=\arctan\frac{w\sin F+(m-1)\sqrt{1-w^2}\cos F}{w\cos
F-(m-1)\sqrt{1-w^2}\sin F}=
F+\psi
\label{psi-psi}
\end{equation}
instead of $\psi$ defined by \eqref{psi}.

Consider now the following moving point of observation:
\begin{equation}
|n|=w(t-N).
\label{as2-n}
\end{equation}
In the latter case 
\begin{equation}
A^{\mathrm{pass}}(\Omega)=\frac{(m-1)\Omega \EXP{\I N(1-w) \arccos \frac{2-\Omega^2}2} }{-(4-\Omega^2)+\I(m-1)\Omega\sqrt{4-\Omega^2}},
\end{equation}
and asymptotics for 
$(\VFF_n^N)^{\mathrm{pass}}$ 
has the form 
\eqref{asymp-J_2-1}, wherein 
%
\begin{equation}
F_0=N(1-w)
.
\label{F=N(1-w)}
\end{equation}

\begin{remark}  
Asymptotics~\eqref{asymp-J_2} can be formally obtained by substituting 
\begin{equation}
F_0=0
\label{F=0}
\end{equation}
into Eqs.~\eqref{asymp-J_2-1}, \eqref{F=F0}.
\end{remark}

\begin{figure}[htb]  
\centering\includegraphics[width=0.8\textwidth]{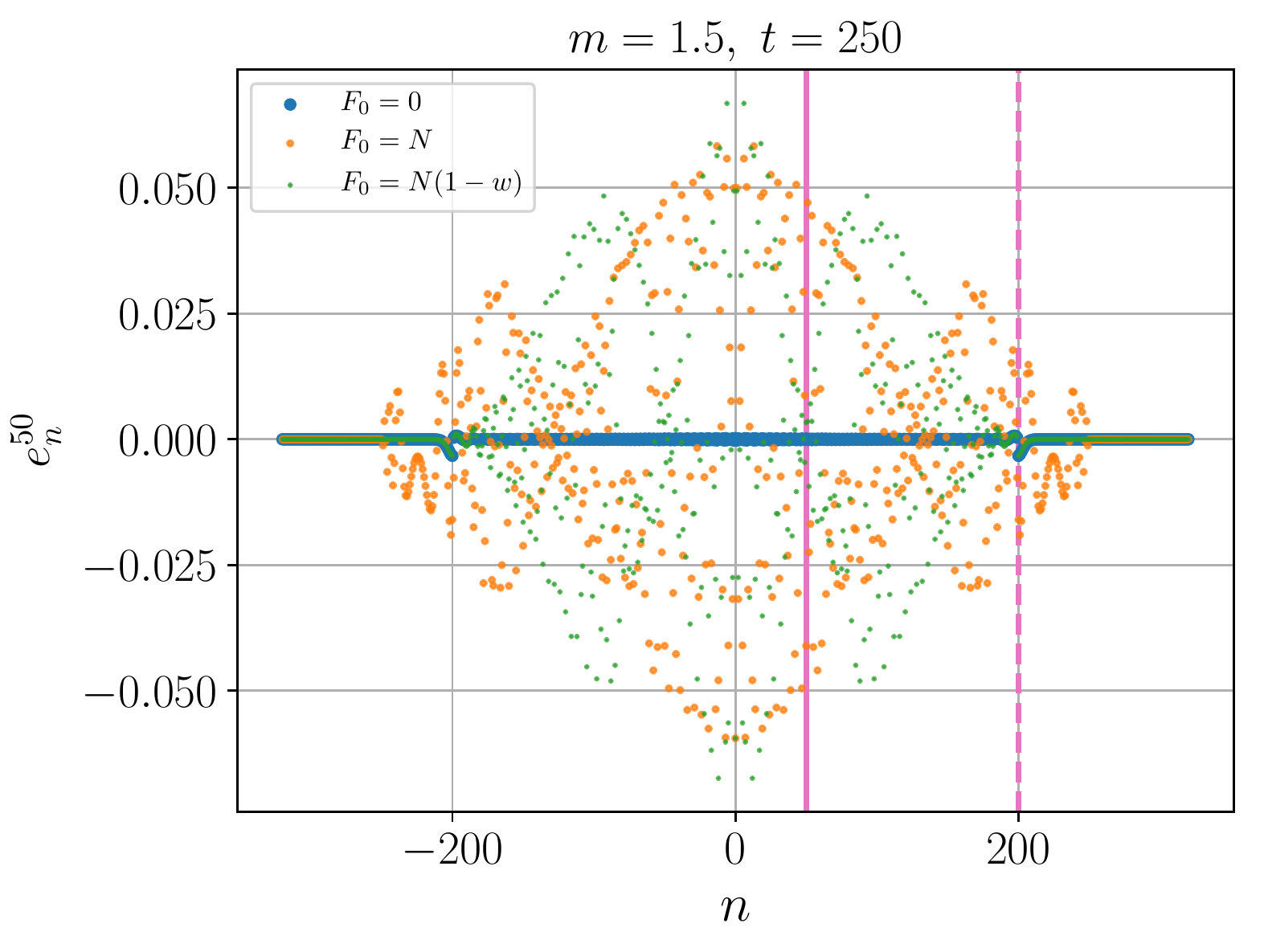}
\caption{Error $e_n^N$ versus $n$. The source position
is indicated
by the vertical magenta solid line. 
The right leading reflected wave-front is indicated
by the vertical magenta dashed line}
\label{error_dif.pdf}
\end{figure}

All asymptotics in the form of Eq.~\eqref{asymp-J_2-1} with various $F_0$
defined by 
\eqref{F=N},
\eqref{F=N(1-w)},
\eqref{F=0}
are
formally correct. {According to 
Eqs.~\eqref{F=F0},
\eqref{psi-psi},
for various $F_0$ the right-hand side of
Eq.~\eqref{asymp-J_2-1}
has the same amplitude but different phases. Moreover, formulae 
\eqref{as1-n},
\eqref{as2-n}, or
\eqref{wt_for-n<0}
that we
use to return to the variables $n$ and $t$ from $w$ and $t$ are also different.}
Therefore, 
the 
applicability of the corresponding asymptotics as an approximate solution in
terms of $n$ and $t$
can also be different. To check which approach is better, we calculate the
absolute error 
\begin{equation}
e_n^N\=
(\VF_n^N)_{\mathrm{num}}-V_{n-N}-(\VFF_n^N)^{\mathrm{pass}}_{\mathrm{approx}}
\end{equation}
using various asymptotics
\eqref{asymp-J_2-1},
\eqref{F=F0}
with $F_0$ defined by 
\eqref{F=N},
\eqref{F=N(1-w)}
or
\eqref{F=0}.
Here 
$(\VF_n^N)_{\mathrm{num}}$ are values for the particle velocities found
numerically, $V_{n-N}$ is given by exact formula \eqref{Sro-bessel},
$(\VF_n^N)^{\mathrm{pass}}_{\mathrm{approx}}$ are found by 
Eq.~\eqref{asymp-J_2-1} wherein $w$ is found in accordance with 
the corresponding formula from set \eqref{as1-n}, \eqref{as2-n},
\eqref{wt_for-n<0}. The plot for the error $e_n^N$ is presented in
Fig.~\ref{error_dif.pdf}. One can see that the choice of $F_0$ in the form of 
Eq.~\eqref{F=0} gives the best result, whereas asymptotics with $F_0$ in the
form of 
\eqref{F=N},
\eqref{F=N(1-w)} are practically inapplicable as an approximate solution in
terms of variables $n$ and $t$.

%

%

%
%


\section{Fixed position asymptotics for $n\leq0$: the amplitude expansion near the cut-off frequency in the pass-band}
\label{app-pass}

Take $n\leq0$. 
For $C_n^N$ defined by Eq.~\eqref{C-def}, one has
\begin{gather}
C_n^N(\Omega)=
\big(A(\Omega)+
B^{}(\Omega)\big)\E,
\label{Cn-dex-ex}
\\
A^{}(\Omega)\E=\Omega \breve {\mathcal G}_n^N,
\qquad
B^{}(\Omega)\E=\Omega G_{n-N}.
\label{ABEn-dex-ex}
\end{gather}
In the pass-band, 
$A^{}(\Omega)$ is defined by Eq.~\eqref{A-pass-upright-def},
\begin{gather}
B(\Omega)=
-\frac{1}
{\I \sqrt{4-\Omega^2}},
\\
\E={\mathrm e}^{\I (\k) \arccos\frac{2-\Omega^2}2},
\end{gather}
see Eqs.~\eqref{G-def},
\eqref{Green-function0-lower},
\eqref{wn-pass},
\eqref{V-1whole}--\eqref{bVNNPass-Stop-def},
\eqref{I-pass-left-pre}. 
For $\Omega\to2-0$, one can obtain the following asymptotic expansions:
\begin{gather}
A(\Omega)=
\frac{-\I}{2 \sqrt{2-\Omega}}
-
\frac{1}{2 (m-1)}
+
\frac{\I(-m^2+2 m+7) \sqrt{2-\Omega}}{16 (m-1)^2}
+O(2-\Omega),
\\
B(\Omega)=
\frac{\I }{2 \sqrt{2-\Omega} }+\frac{\I \sqrt{2-\Omega}}{16}
+O\big((2-\Omega)^{3/2}\big),
\\
\E=(-1)^\k-2 \I (\k) (-1)^\k \sqrt{2-\Omega}+O(2-\Omega).
\end{gather}


Now, one gets
\begin{gather}
A(\Omega)+
B(\Omega)
=
-\frac{1}{2 (m-1)}
+\frac{\I\sqrt{2-\Omega}}{2 (m-1)^2}
+O(2-\Omega),
\\
C_n^N(\Omega)
=
-\frac{(-1)^\k}{2 (m-1)}
+\frac{\I(-1)^\k  \big(1+2(\k)(m-1)\big)\sqrt{2-\Omega}}{2 (m-1)^2}+O(2-\Omega).
\label{Cpass-expansion}
\end{gather}

\section{Fixed position asymptotics for $n\leq0$: the amplitude expansion near
the cut-off frequency in the stop-band}
\label{app-stop}

\allowdisplaybreaks
In the stop-band, we again have Eqs.~\eqref{Cn-dex-ex}, \eqref{ABEn-dex-ex},
wherein
%
\begin{gather}
A(\Omega)=
\frac{\Omega^3(m-1)
}
{\Big(-\Omega^2+2\EXP{-\arccosh\frac{\Omega^2-2}{2}}+2\Big)
\Big(-m\Omega^2+ 2\EXP{- \arccosh\frac{\Omega^2-2}{2}}+2\Big)}
,
\\
B(\Omega)=
\frac{\Omega
}{-\Omega^2+2\EXP{-\arccosh\frac{\Omega^2-2}{2}}+2}
,
\\
\Es=
(-1)^{\k}\EXP{-(\k) \arccosh\frac{\Omega^2-2}{2}},
\end{gather}
see Eqs.~\eqref{G-def},
\eqref{Green-function0-upper},
\eqref{wn-stop},
\eqref{V-1whole}--\eqref{bVNNPass-Stop-def},
\eqref{stop-con}.
For $\Omega\to2+0$, one can obtain the following asymptotic expansions:
\begin{gather}
A(\Omega)=
\frac{1}{2 \sqrt{\Omega-2}}
-\frac{1}{2 (m-1)}
-\frac{\left(m^2-2 m-7\right)\sqrt{\Omega-2}}{16 (m-1)^2}
+O(\Omega-2),
\\
B(\Omega)
=
-\frac{1}{2 \sqrt{\Omega-2}}
+\frac{\sqrt{\Omega-2}}{16}  
+O\big((\Omega-2)^{3/2}\big),
\\
\Es
=(-1)^\k\big(
1
-2 (\k)\sqrt{\Omega-2}+O(\Omega-2)\big)
.
\end{gather}
%

Now, one gets
\begin{gather}
A(\Omega)+
B(\Omega)
=
-\frac{1}{2 (m-1)}+\frac{\sqrt{\Omega-2}}{2 (m-1)^2}
+O(\Omega-2),
\\
C_n^N(\Omega)
=
-\frac{(-1)^\k}{2 (m-1)}+\frac{(-1)^\k  \big(1+2(\k)(m-1)\big)\sqrt{\Omega-2}}{2 (m-1)^2}
+O(\Omega-2).
\label{Cstop-expansion}
\end{gather}

\bibliographystyle{plainnat}
\bibliography{bib/impurity,bib/serge-gost,bib/thermo,bib/math,bib/mode-trans,bib/discrete,bib/graphene,bib/mode,bib/embedded}

\end{document}